\documentclass[a4paper,fleqn,usenatbib]{mnras}

\usepackage{newtxtext,newtxmath}

\usepackage[T1]{fontenc}
\usepackage{ae,aecompl}

\usepackage{graphicx}	
\usepackage{amsmath}	
\usepackage{amssymb}	
\usepackage{nicefrac}   

\graphicspath{ {figures/} }

\newcommand{\p}{\partial}

\newcommand{\M}{{\cal M}}

\newcommand{\rsh}{{r_{\rm sh}}}

\newcommand{\Msol}{\ensuremath{\rm{M}_{\odot}}}
\newcommand{\mns}{M_{\rm{NS}}}

\newcommand{\chicrit}{\chi_{\rm crit}}
\newcommand{\wbv}{\omega_{\rm BV}}
\newcommand{\tadv}{t_{\rm adv}}
\newcommand{\tconv}{t_{\rm buoy}}
\newcommand{\e}{{\rm e}}

\newcommand{\Mc}{{{\cal M}^2}}

\title[The non-linear onset of convection]{The non-linear onset of neutrino-driven convection in two and three-dimensional core-collapse supernovae}

\author[R\'emi Kazeroni et al.]{
R\'emi Kazeroni,$^{1,2}$\thanks{E-mail: kazeroni@MPA-Garching.MPG.DE}
Brendan K. Krueger,$^{2,3}$
J\'er\^ome Guilet,$^{2,1}$
Thierry Foglizzo$^{2}$
\newauthor
and Daniel Pomar\`ede$^{4}$
\\
$^{1}$Max-Planck-Institut f\"ur Astrophysik, Karl-Schwarzschild-Str. 1, D-85748 Garching, Germany \\
$^{2}$Laboratoire AIM, CEA/DRF-CNRS-Universit\'e Paris Diderot, IRFU/D\'epartement d'Astrophysique, CEA-Saclay F-91191, France \\
$^{3}$Eulerian Codes (XCP-2), Los Alamos National Laboratory, Los Alamos, NM 87545, USA \\
$^{4}$Institut de Recherche sur les Lois Fondamentales de l'Univers, CEA, Universit\'e Paris-Saclay, 91191 Gif-sur-Yvette, France
}

\date{Accepted 2018 June 28. Received 2018 June 15; in original form 2018 February 22}

\pubyear{2018}

\begin{document}
\label{firstpage}
\pagerange{\pageref{firstpage}--\pageref{lastpage}}
\maketitle

\begin{abstract}
 A toy model of the post-shock region of core-collapse supernovae is used to study the non-linear development of turbulent motions driven by convection in the presence of advection. Our numerical simulations indicate that buoyant perturbations of density are able to trigger self-sustained convection only when the instability is not linearly stabilized by advection. Large amplitude perturbations produced by strong shock oscillations or combustion inhomogeneities before the collapse of the progenitor are efficiently shredded through phase mixing and generate a turbulent cascade. Our model enables us to investigate several physical arguments that had been proposed to explain the impact of the dimensionality on the onset of explosions in global simulations of core-collapse supernovae. Three-dimensional (3D) simulations are found to lead to higher entropy values than two-dimensional (2D) ones.  We attribute this to greater turbulent mixing and dissipation of the kinetic energy into heat in 3D. Our results show that the increase of entropy is enhanced with finer numerical resolution and larger perturbation amplitude.
\end{abstract}




\section{Introduction}
\label{sec:intro}

During the first second after the core of a massive star collapses into a proto-neutron star, the multidimensional hydrodynamics of the innermost region of few hundreds of kilometres plays a crucial role in driving an explosion of the star, known as a core-collapse supernova (CCSN). The delayed neutrino mechanism \citep{bethe85} represents the most favoured mechanism to explain a large majority of CCSNe (see \citealt{foglizzo15, janka16, mueller16b} for recent reviews). The shock wave formed at core bounce turns into an accretion shock and stalls in the iron core at a radius of about $\rm{150\,km}$ due to energy loss induced by neutrino emission and dissociation of heavy nuclei. Neutrinos carry away most of the energy ($\sim10^{53}\,\rm{ergs}$) released in the gravitational contraction of the core to a proto-neutron star. A small fraction of the neutrino flux is absorbed in the gain layer, a region just below the stalled shock wave where neutrino absorption dominates neutrino emission. A successful explosion is powered and a neutron star is left behind the CCSN if the energy deposition is efficient enough to revive the stalled shock wave, preventing the formation of a black hole \citep{oconnor11}. 

Self-consistent numerical simulations of CCSNe including, among others, detailed neutrino transport and microphysics, three-dimensional hydrodynamics and general relativity still remain computationally challenging. Approximations in treatment of the physics and the numerics, such as a reduction of the dimensionality, are required to gain insight in the CCSN problem.
Numerical calculations performed in 1D demonstrated that spherical symmetry prevents the onset of explosions \citep{liebendoerfer01} except for the lowest CCSN progenitor masses \citep{kitaura06}. A successful explosion relies on the decisive action of multidimensional hydrodynamical instabilities which break the spherical symmetry of the collapse and generate vigorous non-radial motions to large spatial scales. This increases the advection time of matter through the gain region, enhances the heating efficiency and creates more favourable conditions to initiate a CCSN (e.g. \citealt{murphy08}).
Axisymmetric simulations (2D) with various approximations established that two distinct instabilities can govern the dynamics of the gain region prior to the shock revival \citep{mueller12, fernandez14}. These results were later confirmed in 3D \citep{takiwaki14, melson15a, melson15b, lentz15, roberts16}. Neutrino-driven convection was seen in some of the earliest multidimensional CCSNe simulations \citep{herant92, herant94, burrows95, janka96}. The negative entropy gradient induced by neutrino heating can be unstable to buoyancy and convective overturns develop in the gain region, characterized by intermediate spatial scales $l\sim 5-6$. The second instability is the Standing Accretion Shock Instability (SASI) \citep{blondin03} mediated by an advective-acoustic cycle \citep{foglizzo07,foglizzo09,guilet12}. It generates global oscillatory motions of the shock wave associated with the largest spatial scales $l\sim1-2$. 

Large scale asymmetries in the collapsing core are supported by some observational evidence. 
The development of early asymmetries is suggested by the spectropolarimetric measurement of a type II-P supernova during the nebular phase of the explosion \citep{leonard06}.
High kicks of young neutron stars \citep{arzoumanian02} could result from a large scale asymmetry $l=1$ present at the onset of the explosion \citep{scheck04, scheck06}. The one-sided spatial distribution of ${}^{44}\rm{Ti}$ observed in Cassiopea A \citep{grefenstette14} seems to confirm numerical calculations that predict large scale asymmetries seeded by hydrodynamical instabilities \citep{wongwathanarat13,wongwathanarat17}. Convection and SASI could leave different imprints on the gravitational waves \citep{mueller13, kuroda16, andresen17} and the neutrino signals \citep{lund12, tamborra13, tamborra14a} that might be detected for a future nearby CCSN.

A majority of 2D simulations produce explosions that are underenergetic compared to the range of explosion energies expected from observations \citep{mueller15}. It remains unclear whether releasing the axisymmetric constraint reduces the discrepancies, in part because running self-consistent 3D simulations long enough to infer the explosion energy is barely feasible. Moreover, the impact of the dimensionality on the likelihood to explode is still controversial. Models assuming a light-bulb approximation for the neutrino transport reached different conclusions regarding the critical neutrino luminosity required to obtain an explosion. \citet{burrows12,dolence13} found slightly more favourable conditions in 3D while \citet{hanke12} saw almost no difference and \citet{couch13a} obtained delayed explosions in 3D. More accurate neutrino transport schemes  provided less ambiguous results since the shock revival either fails in 3D \citep{hanke13, tamborra14b} or is notably delayed compared to the corresponding 2D simulation \citep{takiwaki14, melson15b, lentz15}. The first 3D simulations that produced more robust explosions associated with a faster growth of the energy \citep{melson15a, mueller15} raised hope that 3D is not necessarily detrimental to CCSNe. Besides, the inclusion of multidimensional initial conditions obtained by simulating shell burning a few minutes prior to collapse \citep{mueller16a} may lead to an explosion energy in the ballpark of observations \citep{mueller17}.

Several physical arguments have been put forward to interpret the discrepancies between 2D and 3D simulations. The turbulent energy cascade acts in different directions in 2D and in 3D \citep{hanke12}. In 2D, the reverse turbulent cascade feeds the largest spatial scales \citep{kraichnan67} and favours the formation of large buoyant bubbles that help to trigger shock revival \citep{fernandez14}. On the contrary, the forward cascade in 3D transfers energy to small scales and induces more dissipation of turbulent kinetic energy. Turbulence created by the instabilities in the post-shock region provides additional pressure support to revive the shock wave. On the one hand, it was proposed that turbulent kinetic energy, related to the largest spatial scales, is artificially overpredicted in 2D simulations \citep{murphy13, couch15} and this could lead to earlier explosions. On the other hand, turbulent dissipation to small scales could play a positive role in 3D \citep{mabanta18} but this has to be tested in multidimensional simulations. It was also suggested that 2D buoyant bubbles undergo a weaker drag force due to their particular shapes and drive earlier explosions \citep{couch13a}. In their simulations of low-mass progenitors dominated by the convective instability, \citet{melson15a} and \citet{mueller15} observed that positive 3D effects could come into play after shock revival. Greater fragmentation and mixing to small scales at the edge of downflows are able to decelerate them, keeping more matter in the gain region and fostering explosions in 3D. Whether these effects are generic and play a role in more massive progenitors is not yet known.

Using analytical calculations, \citet{foglizzo06} showed that convection can be stabilized by a fast advection through the gain region. Therefore a negative entropy gradient is not a sufficient instability criterion because the exchange of matter induced by buoyancy should take place before the perturbation leaves the gain region. The linear analysis of \citet{foglizzo06} is strictly speaking only valid for perturbations of small amplitudes. It has been proposed that strong perturbations could possibly trigger convection in situations where it is linearly stabilized by advection \citep{scheck08,fernandez14}. SASI acts as a source of perturbations because the instability pushes the shock outward and creates entropy-vorticity waves which can feed secondary convection \citep{scheck08, guilet10, cardall15, summa16}. Progenitor inhomogeneities produced by convective shell burning create multidimensional initial conditions which can ease the shock revival under certain circumstances \citep{couch13b, couch15, mueller15a}. It was shown that this mechanism could lead to the formation of large scale structures ($l\sim2-4$) which can turn a failed explosion into a successful one \citep{mueller16a,mueller17}. Lastly, the noise related to some numerical features such as the grid geometry may also seed convection. 
A Cartesian grid with several refinement levels towards the centre of the domain generates greater perturbations than a spherical grid. 
Such perturbations alone could impact the dynamics to the point of selecting the dominant instability during the whole simulation. 
A $27\,\Msol$ progenitor was found to be dominated by convection \citep{ott13} or by SASI after overcoming the grid artefacts \citep{abdikamalov15}. 

These numerous works have explored different aspects of the multidimensional dynamics and yet several key questions remain uncertain. What is the main physical effect that drive multidimensional simulations closer to explosions? Several effects have been put forward such as a longer residency timescale in the gain layer \citep{murphy08}, the turbulent pressure pushing the shock \citep{couch15} or heating by dissipation of kinetic energy \citep{mabanta18} but their relative importance remain to be determined. What are the main differences between 2D and 3D? What is the role of perturbations resulting from the pre-collapse dynamics or seeded by numerical errors? Can they trigger self-sustained convection in the regime where advection linearly stabilizes the instability? Addressing these issues and disentangling the different effects in global CCSN simulations is challenging, such that a simplified setup is an important complementary approach. The simplicity of such models is an asset not only for the physical interpretation but also to allow an easier exploration of the parameter space and to study the impact of numerical resolution.

In this study, we investigate the competition between advection and convection in its simplest form. An idealized description of the advection of matter through the gain layer during the stalled shock phase of a CCSN is employed. Numerical simulations of the toy model enable us to assess the robustness of linear and non-linear instability criteria. We show that a strong perturbation is not a sufficient condition to lead to fully developed convection and then discuss several regimes of the instability. 
Various interpretations of the impact of dimensionality on convection are revisited in the light of our model that reproduces the classical properties of turbulence induced by the instability. Our simulations show an excess of heating in 3D due to the turbulent dissipation of kinetic energy. We identify the hydrodynamical processes that could explain this phenomenon based on the particular dynamics at small spatial scales. We discuss the influence of the numerical resolution and show that a greater impact on the dynamics is expected when large amplitude perturbations come into play.

The rest of the paper is organized as follows. The physical model and the numerical methods are detailed in Section \ref{sec:model}. Predictions regarding the linear and non-linear onsets of the convective instability are tested in Section \ref{sec:regimes}. An overview of the main discrepancies between the 2D and the 3D dynamics is presented in Section \ref{sec:quali}. A quantitative analysis of the properties of turbulence is proposed in Section \ref{sec:quanti} to pinpoint the  effects that could explain the main differences between our 2D and 3D simulations. The influence of the numerical resolution is discussed in Section \ref{sec:reso}. The limitations of our model and the consequences of our results for self-consistent simulations are addressed in Section \ref{sec:discussion}. We summarize our findings and conclude in Section \ref{sec:conclusion}.

\section{Physical and numerical setup}
\label{sec:model}

Our approach focuses on an idealized description of the accretion flow below the stalled shock wave of a CCSN and includes only the minimal ingredients leading to the convective instability. This phase is modelled by a stationary flow crossing a source layer which represents a fraction of a realistic gain layer and contains gravity and neutrino heating. The shock wave and the cooling layer surrounding the proto-neutron star are absent from our model in order to study convection in its simplest form and preclude any feedback that would interact with the dynamics of the gain region.

\subsection{Stationary flow}
\label{subsec:flow}

The stationary flow considered in this study is similar to the one studied by \citet{foglizzo06}. It is planar along the vertical direction (z). The flow is modelled as a perfect gas with an adiabatic index $\gamma=4/3$. This is a suitable approximation of the equation of state in the post-shock region where the pressure is dominated by relativistic electrons, photons, and electron-positron pairs \citep{janka01}.

In our model, the gain region contains the effects of heating and gravity and is located in the central part of the domain.
Upstream and downstream from this layer, gravity and heating are turned off and the stationary flow is thus uniform. The two external layers are used to place the boundaries far enough in the vertical direction in order to minimize the impact of numerical reflections on the instability. To avoid any discontinuity at the horizontal edges of the source layer, gravity and heating are smoothed out by a linear ramp function $\Psi\left(z\right)$:
\begin{equation}
\label{eq:shape}
 \Psi\left(z\right) \equiv 
 \begin{cases}
    1 &\text{ if } \left| z \right| < \frac{H}{2}, \\
    2\left(1-\frac{z}{H}\right) &\text{ if } \frac{H}{2} \leq \left| z \right| < H, \\
    0 &\text{ if } H \leq \left| z \right|,
 \end{cases}
\end{equation}
where $2H$ corresponds to the height of the gain layer. To ease the comparison with more realistic models, $H$ is set to 50 km. Variations of $H$ due to shock oscillations and the recession of the gain radius over time are neglected for the sake of simplicity. The numerical domain is delimited in the horizontal directions by $-150\,\rm{km} \leq x,y \leq 150\,\rm{km}$ and in the vertical direction by $-450\,\rm{km} \leq z \leq 450\,\rm{km}$. The horizontal extent of our domain covers about one third of the angular size of a gain region for a shock radius of 150 km. The gain layer is delimited vertically by $-50\,\rm{km} \leq z \leq 50\,\rm{km}$.

The gravitational potential is defined as:
\begin{equation}
\label{eq:Hgrav}
\nabla\Phi \equiv K_G \left(\frac{c^2_{\rm{up}}}{H}\right) \Psi\left(z\right),
\end{equation}
where $K_G$ is a dimensionless parameter that quantifies the intensity of gravity, $c$ is the sound speed and the subscript ``up'' refers to quantities upstream from the source layer.
The energy deposition by neutrinos per unit of time in the source layer is modelled by a heating function of the form:
\begin{equation}
\label{eq:Hheat}
 \mathcal{L} \equiv K_H \left(\frac{\rho_{\rm{up}} \mathcal{M}_{\rm{up}} c^3_{\rm{up}}}{\gamma H} \right) \left(\frac{\rho}{\rho_{\rm{up}}}\right) \Psi\left(z\right),
\end{equation}
where $\rho$ corresponds to the density and $\mathcal{M}$ to the Mach number defined as: $\mathcal{M} \equiv \left| v_z\right|/c$ with $v$ being the fluid velocity.
The dimensionless numbers $K_G$, $K_H$ and $\mathcal{M}_{\rm{up}}$ represent the free parameters of our model. In the numerical simulations of the model, we consider only variations of $K_H$, the normalization of the heating function. This parameter controls whether convection is linearly unstable or stabilized by advection (see Sect. \ref{subsec:analysis} and appendix \ref{sec:appendixB}).

\begin{figure}
\centering
	\includegraphics[width=0.9\columnwidth]{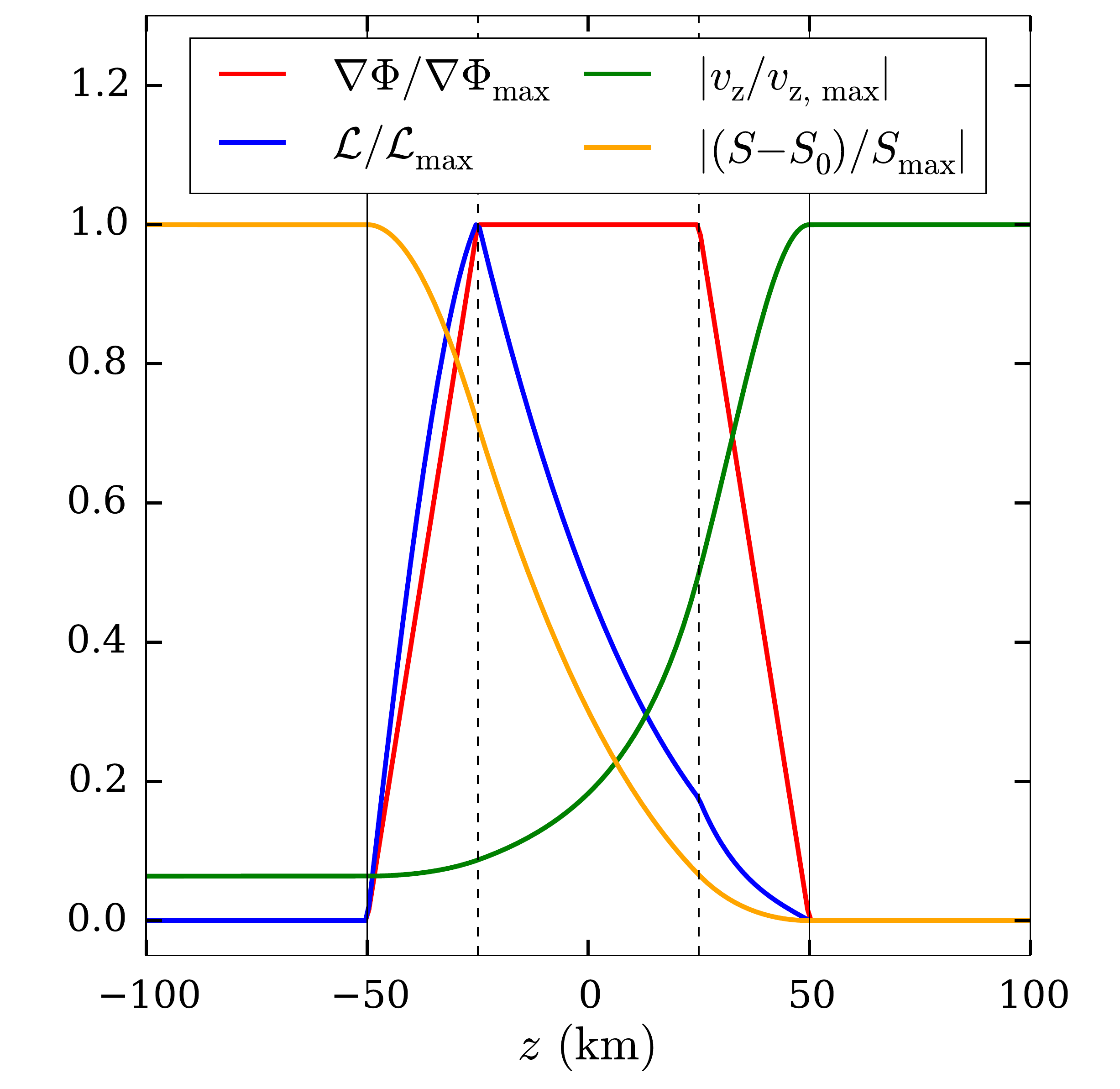}
    \caption{Vertical profiles of gravity (red), vertical velocity (green), heating (blue) and entropy (orange) across the gain region. The quantities are normalized by their maximum values except the entropy which is shown as $\left(S-S_0\right)/S_{\rm max}$ where $S_0$ defines the uniform entropy upstream from the gain region. The upstream flow (positive $z$) is decelerated by the gravity step function and its entropy increases due to heating. The outer edges of the source layer ($\Psi(z)>0$) are delimited by the vertical solid lines and the region where the source terms are at their full intensity ($\Psi(z)=1$) is delimited by the vertical dashed lines.
    }
    \label{fig:init}
\end{figure}

The parameter $\mathcal{M}_{\rm{up}}$ is the equivalent to the post-shock Mach number. Its value depends on the equation of state and the description of the matter. For an ideal gas with an adiabatic index $\gamma=4/3$, it varies from $\mathcal{M}_{\rm{up}} \approx 0.1$ when nuclei are completely dissociated by the shock wave to $\mathcal{M}_{\rm{up}} \approx 0.3$ for an adiabatic shock \citep{fernandez09b}. In this study, we consider the case $\mathcal{M}_{\rm{up}} = 0.3$.

An estimate of the normalization of the gravitational potential can be made using $H$ and $c_{\rm{up}}$. The latter can be computed as
\begin{equation}
 \label{eq:cup}
 c_{\rm up}^2 = \frac{v^2_{\rm up}}{\mathcal{M}^2_{\rm{up}}} = \frac{v^2_{\rm ff}}{\mathcal{M}^2_{\rm{up}} \kappa^2} = \frac{2G\mns}{\rsh}\frac{1}{\mathcal{M}^2_{\rm{up}} \kappa^2},
\end{equation}
where $\kappa$ corresponds to the compression factor across the shock and $v_{\rm ff}$ to the free-fall velocity at the shock radius. The parameter $\kappa$ ranges from about 5 for an adiabatic shock to about 10 for a strong shock that fully dissociates iron nuclei. 
Using equations (\ref{eq:Hgrav}) and (\ref{eq:cup}), one obtains:
\begin{equation}
 \label{eq:Kg}
 K_G = \nabla \Phi\frac{H}{c_{\rm up}^2} = \left(\frac{\rsh}{r}\right)^2\frac{H}{\rsh}\frac{\mathcal{M}^2_{\rm{up}} \kappa^2}{2}.
\end{equation}
During the stalled shock phase, one typically finds that $\rsh/r \sim 1.4-3$ and $H/\rsh \sim 0.3-0.7$ (e.g. \citealt{marek09, melson15a}). Thus, the normalization of the gravitational potential is such that $K_G \sim 0.6-5.4$ for an adiabatic shock and $K_G \sim 0.3-2.7$ for a strong shock. In the following, we restrict ourselves to $K_G =3$. 

The structure of the stationary flow can be obtained by solving the following equations:
\begin{align}
 \frac{\partial \rho v_z}{\partial z} &= 0, \label{eq:conv_stat1} \\ 
 \frac{\partial}{\partial z} \left( \frac{v^2}{2} + \frac{c^2}{\gamma-1}+ \Phi \right) &= \frac{\mathcal{L}}{\rho v_z}, \label{eq:conv_stat2} \\ 
 \frac{\partial S}{\partial z} &= \frac{\mathcal{L}}{P v_z}, \label{eq:conv_stat3}
\end{align}
where $S$ defines the dimensionless entropy
\begin{equation}
 S \equiv \frac{1}{\gamma-1}\log\left[\left(\frac{P}{P_{\rm{up}}}\right)\left(\frac{\rho_{\rm{up}}}{\rho}\right)^{\gamma}\right],
\end{equation}
with $P$ being the pressure. When crossing the gain layer, the subsonic flow is decelerated by the gravity function and its entropy increases due to the heating function which is proportional to the density (Fig. \ref{fig:init}).

\subsection{Linear analysis}
\label{subsec:analysis}

The presence of a negative entropy gradient in the gain region is not sufficient to drive convection because the advection timescale through the gain region is finite \citep{foglizzo06}. The flow is stable to convection if the kinetic energy of the accretion flow is high enough to overcome the gravitational potential energy liberated in the vertical exchange of low and high entropy material. \citet{foglizzo06} proposed a linear criterion to assess whether convection can develop or if it is stabilized by advection. This criterion relies on a dimensionless parameter that compares the local buoyancy timescale ($\tconv$) to the advection timescale ($\tadv$) through the gain region and writes for our setup:
\begin{equation}
 \label{eq:chi}
 \chi \equiv \int_{z=-50\,\rm{km}}^{z=50\,\rm{km}}\: \left|\mathrm{Im}\left(\langle\wbv^2\rangle^{1/2}\right) \frac{dz}{\langle v_z\rangle}\right| \sim \frac{\tadv}{\tconv},
\end{equation}
where $\wbv$ corresponds to the Brunt-V\"{a}is\"{a}l\"{a} frequency defined as:
\begin{equation}
 \label{eq:wbv}
 \wbv \equiv \left(\nabla\Phi\right)^{1/2} \left|\frac{\nabla P}{\gamma P} - \frac{\nabla \rho}{\rho}\right|^{1/2} = \left(\frac{\gamma-1}{\gamma}\nabla\Phi\nabla S\right)^{1/2}.
 \end{equation}
 and $\langle.\rangle$ to the horizontal and temporal average over the gain region. The parameter $\chi$ is computed by using the average entropy gradient $\langle\nabla S\rangle$ in Eq. (\ref{eq:wbv}).
The local buoyancy timescale can be approximated by: $\tconv \sim \wbv^{-1}$ while the advection timescale through the gain region is defined as:
\begin{equation}
 \label{eq:tadvdef}
  \tadv \equiv \int_{z=-50\,\rm{km}}^{z=50\,\rm{km}}\: \frac{dz}{\left|\langle v_z\rangle\right|}.
\end{equation}
In the rest of the paper, the time is normalized by the advection time of the unperturbed flow through the gain region. The latter corresponds to about 20 ms and is almost constant over the range of $K_H$ explored in our study.

The linear analysis of the model is detailed in appendix \ref{sec:appendixB}. Fig. \ref{fig:analysis} demonstrates that if the buoyancy timescale is short enough compared to the advection timescale ($\chi>\chi_{\rm crit}$), the flow is unstable for a range of horizontal wavenumbers $[k_{\rm min},k_{\rm max}]$ in a similar manner as in a shocked flow (Fig.~5 in \citealt{foglizzo06}). 
The absence of a shock in the present setup affects the instability threshold $\chi_{\rm crit}$ and the asymptotic scaling of $k_{\rm min}$ as shown in Fig. \ref{fig:kmax}. This figure includes the stability properties of a flow with a strong shock where energy losses by the dissociation of nuclei is chosen such that the postshock Mach number is the same as in our model. The critical instability threshold is lowered from $\chi_{\rm crit}\sim 3.5$ with a shock to $\chi_{\rm crit}\sim 2.4$ without a shock. The effect of a shock on the stability properties is studied analytically in the low mach number limit in appendix \ref{sec:appendixB}, which confirms that the instability threshold is set by the parameter $\chi \propto (K_HK_G)^{1/2}/{\cal M}$ (Eq.~\ref{eq:chi_params}) and demonstrates that the asymptotic scaling of $k_{\rm max}\propto \chi/H$ is not affected by the shock.

\begin{figure}
\centering
	\includegraphics[width=0.9\columnwidth]{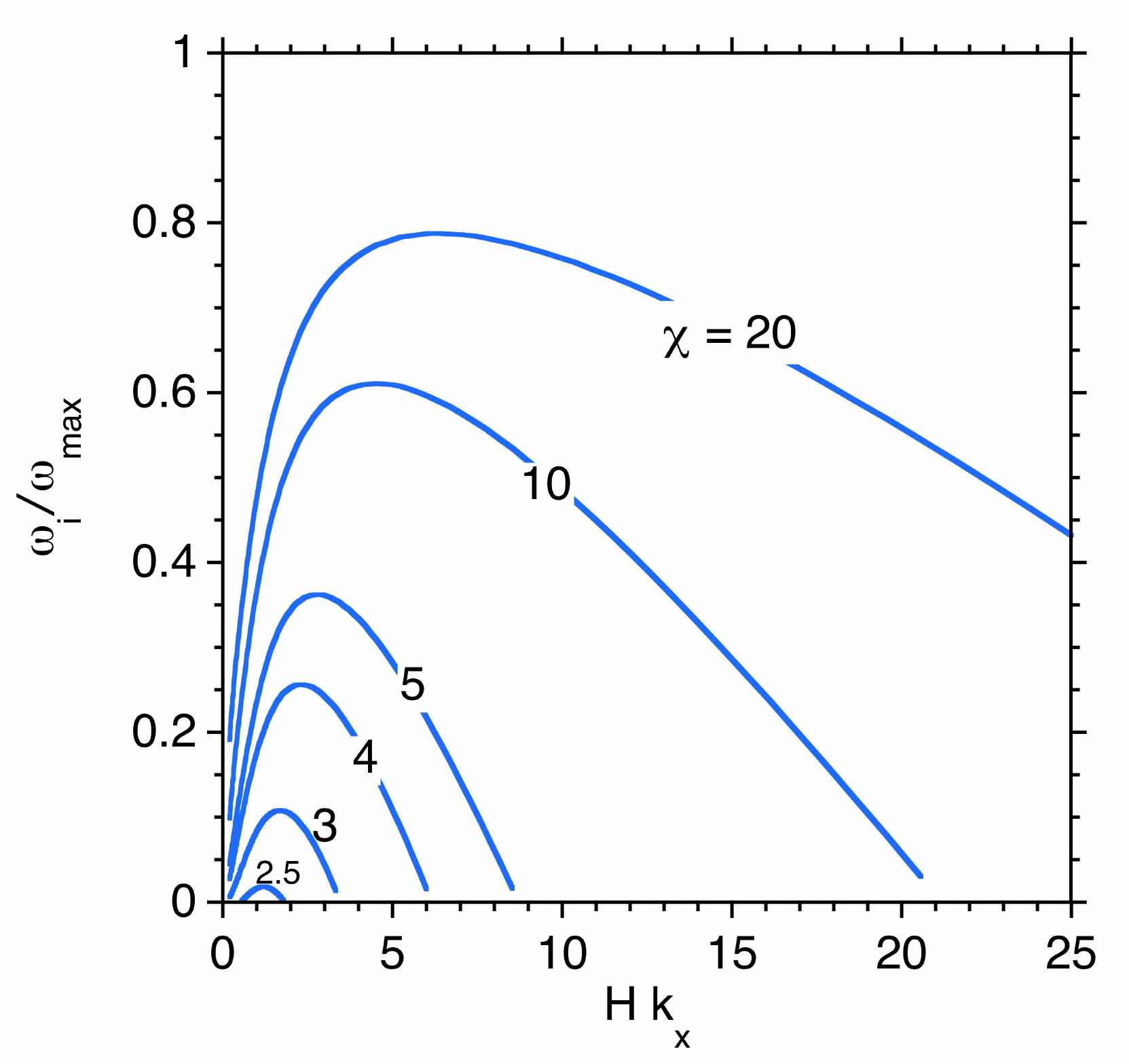}
    \caption{Growth rate of the convective instability as a function of the horizontal wavenumber for $K_G=3$ and $\mathcal{M}_{\rm up}=0.3$. The growth rates are normalized to the maximum value of the Brunt-V\"{a}is\"{a}l\"{a} frequency noted $\omega_{\rm max}$. The ratio $\chi$ of the advective and convective timescales is indicated on each curve. The convective instability disappears for $\chi<\chi_{\rm crit}\sim 2.4$.
    }
    \label{fig:analysis}
\end{figure}

\begin{figure}
\centering
	\includegraphics[width=0.9\columnwidth]{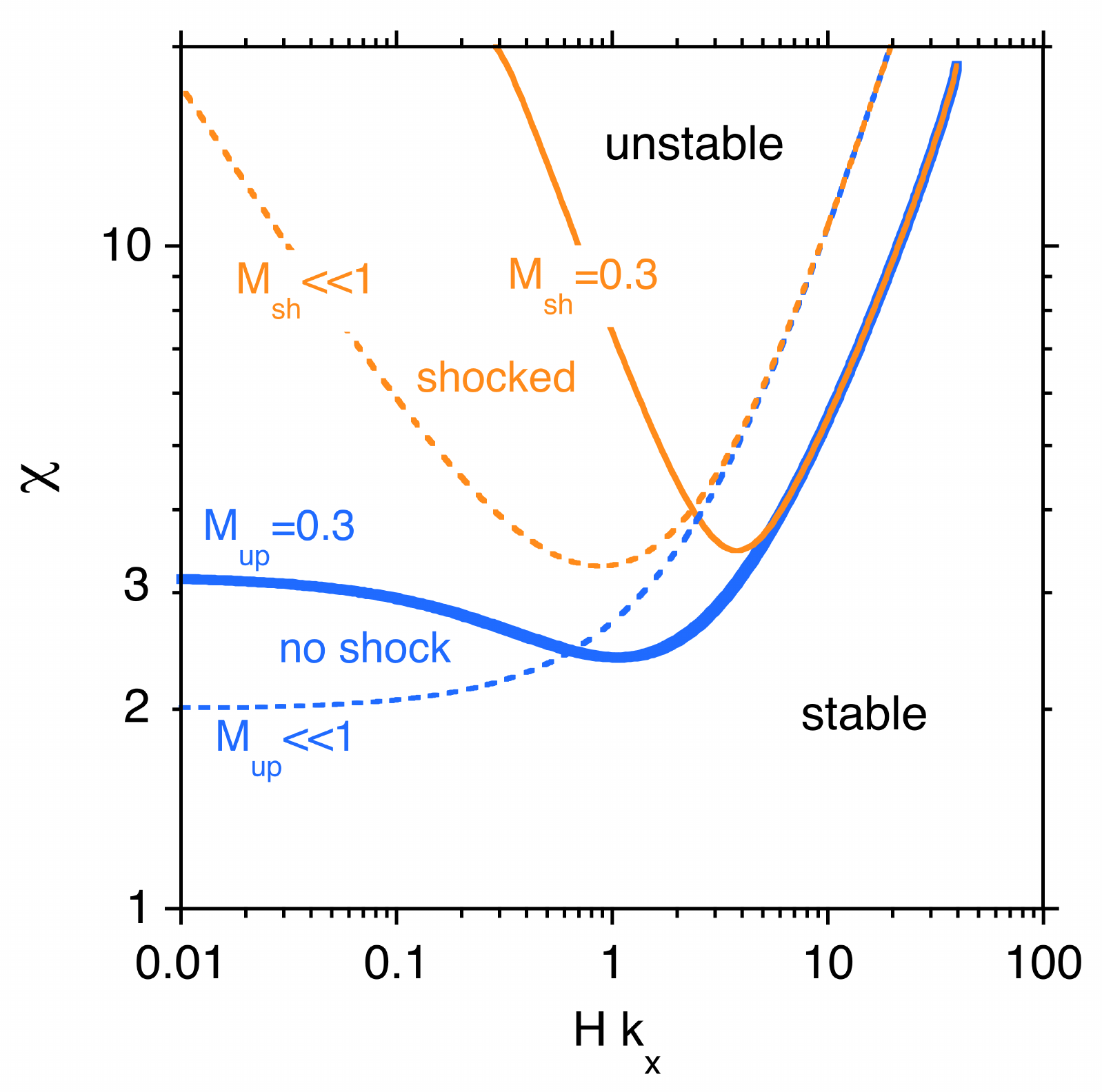}
        \caption{Range of horizontal wavenumbers allowing for the convective instability with and without a shock at the upper boundary. The full lines correspond to an upper Mach number $\M=0.3$ with $K_G=3$. The dashed lines correspond to analytical calculations in the asymptotic limit $K_G\ll1$, $K_H\ll1$, $\M\ll1$ without and with a shock (Eqs. \ref{eq:chi_noshock} and \ref{eq:chi_shock} respectively). The absence of the shock favours the instability at smaller wavenumber and a lower instability threshold $\chicrit\ge2$.
    }
    \label{fig:kmax}
\end{figure}

In the limit $\mathcal{M}\ll1$, the critical instability threshold is lowered from $\chicrit=3.3$ with a shock to $\chicrit=2$ without a shock. Altogether, the convergence study illustrated by Fig. \ref{fig:convergence} suggests that the instability threshold is in the range $2-2.4$ without a shock, and in the range $3.3-3.5$ with a shock. For comparison, a broader range $2.8-3.5$ of critical thresholds was obtained with the combined action of heating, cooling and dissociation below a shock (Fig. 6 in \citealt{foglizzo06}).

\subsection{Initial perturbation}
\label{subsec:perturbations}

\begin{figure}
\centering
	\includegraphics[width=0.65\columnwidth]{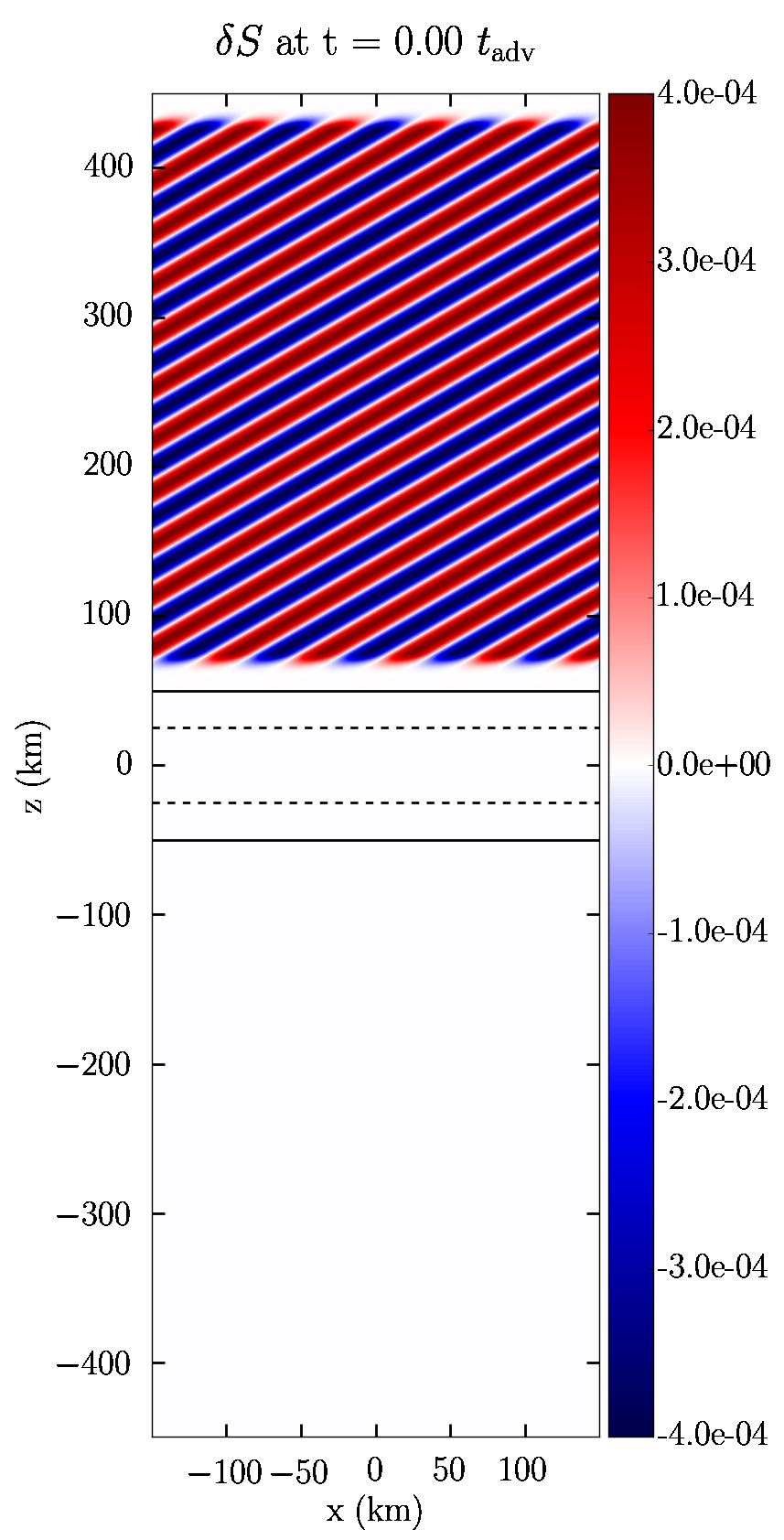}
    \caption{Structure of the perturbation shown in entropy contrast, where the horizontal average value is subtracted from the flow. Outside of the two black solid lines, gravity and heating are turned off ($\Psi(z)=0$), while they are at their full intensity ($\Psi(z)=1$) inside the black dashed lines.
    }
    \label{fig:perturbation}
\end{figure}

The multidimensional dynamics is triggered by a density perturbation at pressure equilibrium which is added to the stationary flow upstream from the source layer (Fig. \ref{fig:perturbation}). The perturbation is initially located between $z=75\,\rm{km}$ and $z=425\,\rm{km}$, thus entirely contained in the upstream region. Its vertical extent is set so that the accretion of the whole perturbation into the gain region corresponds to roughly one advection time of the stationary flow (Eq. \ref{eq:tadvdef}). Note that no perturbations are present at the upper boundary condition ($z=450\,\rm{km}$). In 3D, the perturbation is almost uniform in the transverse direction in order to perform a detailed comparison between the early non-linear regimes in 2D and in 3D. A random noise of amplitude 0.1\% in density is included in each numerical cell to enable the growth of the instability in the third direction.

The perturbation models an entropy wave produced by the advective-acoustic cycle of SASI (Fig. \ref{fig:perturbation}). We aim at studying the coupling between the hydrodynamical instabilities considering a transient excitation of convection by SASI. The absence of a shock wave prevents a possible feedback from the gain region to generate new entropy waves which would continuously feed the instability.

This perturbation could also represent pre-collapse asymmetries originating from combustion inhomogeneities in multi-dimensional progenitors (e.g. \citealt{couch15, mueller16a}) or numerical artefacts such as embedded Cartesian grids refined towards the centre of the domain (e.g. \citealt{ott13}).

Our study focuses on entropy perturbations and ignores vorticity perturbations inherent to the advective-acoustic cycle of SASI \citep{foglizzo07} because those would be unstable to the Kelvin-Helmholtz instability in the uniform flow before reaching the gain layer.
Entropy perturbations are left unchanged upstream from the gain layer and are simply advected until they possibly become unstable due to the action of buoyancy. To restrict the number of parameters, we set the horizontal wavenumber of the perturbation to $m=5$. This corresponds to the most unstable mode for the size of the gain region considered in our study.

\subsection{Numerical simulations}
\label{subsec:simulations}

The RAMSES code \citep{teyssier02, fromang06} is employed to simulate the dynamics on a Cartesian grid. It is a second-order finite volume code which uses the MUSCL-Hancock scheme. We employ the HLLD Riemann solver \citep{miyoshi05} and the monotonized central slope limiter.  
Periodic boundary conditions are employed at the lateral edges of the domain. A constant inflow determined by the stationary flow is used at the top boundary condition. The bottom one consists of a constant outflow where ghost cells are filled with values from the stationary flow. We find that this choice minimizes the reflections at the bottom of the box and only marginally perturbs the constant inflow from the upper boundary. The default numerical resolution is such that $(N_x\times N_z)=(384\times1152)$ in 2D and $(N_x\times N_y\times N_z)=(384\times384\times1152)$ in 3D. The size of the numerical cells is the same in all directions. The impact of the numerical resolution will be addressed in Section \ref{sec:reso}. The gain region is discretized in 128 vertical cells in the z direction and this represents about 60 cells per pressure scale height.

Two main parameters are varied to explore the different regimes of the convective instability and the impact of the dimensionality. The parameter $\chi_0$ corresponds to the initial value of $\chi$ (Fig. \ref{fig:kmax}). It is varied from 0 to 5 to study the regime of linear instability ($\chi>\chi_{\rm crit}$) as well as non-linearly triggered convection ($\chi<\chi_{\rm crit}$). We also vary the amplitude of the perturbation $\delta\rho/\rho$ such that $0.1\%\leq\delta\rho/\rho\leq30\%$. In Section \ref{sec:regimes}, we restrict ourselves to 2D simulations to cover the parameter space ($\chi_0$, $\delta\rho/\rho$) and assess the robustness of linear and non-linear instability criteria. A detailed analysis of the differences between 2D and 3D simulations is presented in Section \ref{sec:quali} based on simulations which illustrate the different instability regimes identified in the next section.

\section{Numerical exploration of the different instability regimes}
\label{sec:regimes}

\subsection{Convective instability triggered non-linearly}
\label{subsec:nonlinear}

A buoyant bubble of density $\rho$ is expected to rise against a surrounding flow of density $\rho_0$, regardless of the value of $\chi$, if the density contrast is such that
\begin{equation}
\label{eq:scheck08}
\delta_{\rm min} \equiv \frac{\left|\rho_0-\rho\right|}{\rho_0} \gtrsim \frac{\langle\left| v_r\right|\rangle_g}{\langle g\rangle_g\tadv} \sim \mathcal{O}\left(1\%\right),
\end{equation}
as proposed by \citet{scheck08}.
This can be understood as a competition between the buoyant acceleration experienced by the bubble and the advection velocity. \citet{fernandez14} proposed an alternate criterion based on the balance between the buoyant force ($\sim V\delta\rho g$) and the drag force ($\sim 1/2 C_D S \rho_0 v^2$) exerted on a bubble:
\begin{equation}
 \label{eq:fernandez14}
 \delta_{\rm min} \equiv \frac{\left|\rho_0-\rho\right|}{\rho_0} \gtrsim \frac{C_D \langle\left| v_r\right|\rangle^2_g}{2l_0\langle g\rangle_g},
\end{equation}
where $C_D$ is the drag coefficient (0.5 for a sphere) and $l_0$ the ratio between the volume $V$ and the cross section $S$ of the bubble.

\begin{figure}
\centering
	\includegraphics[width=0.9\columnwidth]{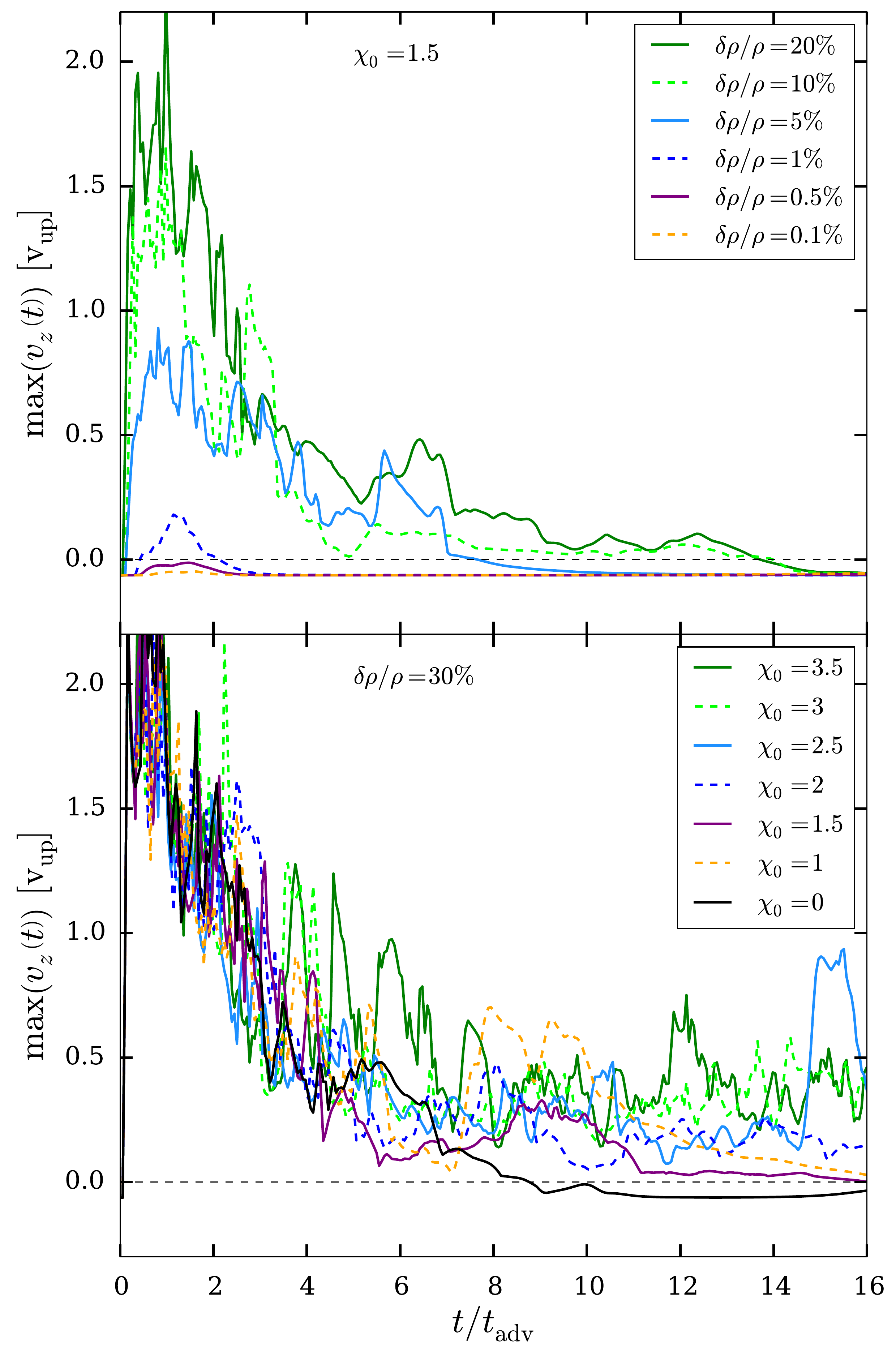}
    \caption{Time evolution of the maximum upward velocity in 2D simulations. The velocities are normalized by the upstream velocity.
    \textit{Top panel:} Set of simulations of linearly stable flows with $\chi_0=1.5$ and different perturbation amplitudes. In all cases, the velocity reaches negative values showing that buoyant motions are suppressed.
    \textit{Bottom panel:} Set of simulations with $\delta\rho/\rho=30\%$ and different values of $\chi_0$. Negative values are close to be reached in all cases where $\chi<\chi_{\rm crit}\sim2.4$ showing that the instability is not self-sustained when triggered in a linearly stable flow.
    }
    \label{fig:amplitude_velo}
\end{figure}

To test these criteria, we perform a set of simulations using a linearly stabilized flow with $\chi_0=1.5$ and different perturbation amplitudes. 
The criteria (\ref{eq:scheck08}) and (\ref{eq:fernandez14}) respectively give thresholds of $0.5\%$ and $0.8\%$, which are in good agreement with our numerical simulations (Fig. \ref{fig:amplitude_velo}, top panel).
However, our results show that perturbation amplitudes slightly above the thresholds (\ref{eq:scheck08}) and (\ref{eq:fernandez14}) are not strong enough to trigger convective overturns but only temporary buoyant motions. For such perturbation amplitudes, the maximum vertical velocity in the gain region becomes negative after a few advection times showing that ascending motions are completely suppressed. 
Our simulations indicate that the criteria used to estimate the rise of a buoyant bubble are not sufficient conditions to trigger turbulent convection.

\begin{figure*}
\centering
	\includegraphics[width=0.7\columnwidth]{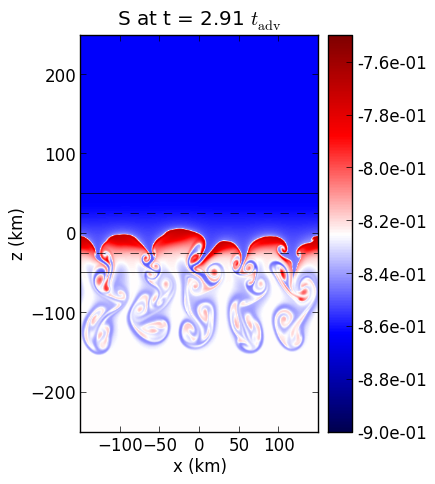}
	\includegraphics[width=0.7\columnwidth]{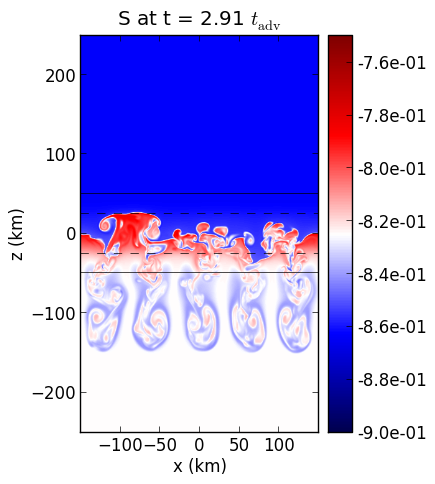}
	\includegraphics[width=0.7\columnwidth]{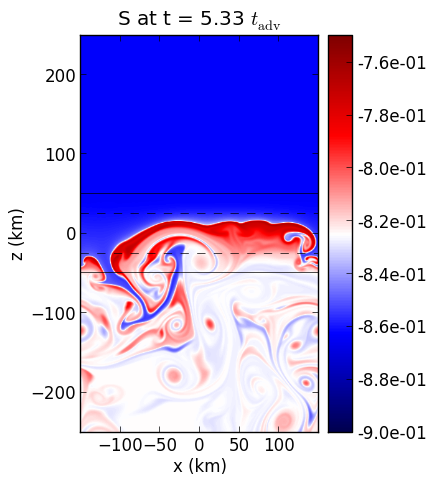}
	\includegraphics[width=0.7\columnwidth]{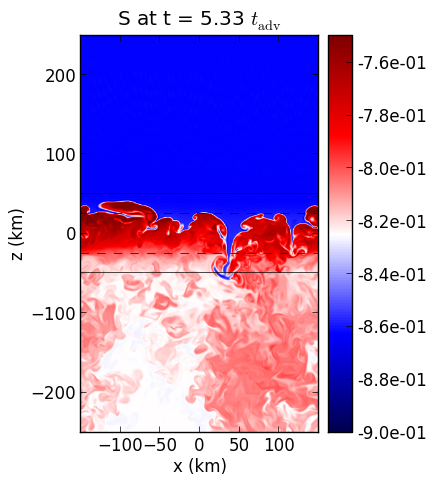}
    \caption{Snapshots of entropy for models with $\chi_0=5$ and $\delta\rho/\rho=0.1\%$ after 2.91 (top) and 5.33 (bottom) advection timescales in 2D (left) and in 3D in the vertical slice $y=0$ (right). The horizontal black lines are defined in Fig. \ref{fig:perturbation}. An animated version of this figure is available in the online journal.}
    \label{fig:snapshots}
\end{figure*}

\subsection{A self-sustained instability?}
\label{subsec:sustained}

We now focus on the dynamics in the gain region at later times. In the set of simulations described in the previous section, we observe that even for the largest perturbation amplitude considered, the instability is completely damped after 15 advection timescales (about 300 ms) since ascending motions no longer exist (Fig. \ref{fig:amplitude_velo}, top panel). In each simulation, the accretion of the initial perturbation into the gain layer lasted less than the first advection timescale. This set of simulations shows that if a large perturbation triggers convection in a linearly stabilized flow, the instability does not reach a permanent turbulent regime and convection is eventually suppressed. The damping timescale increases with stronger perturbations but the flow always adjusts itself to a linearly stable flow (sub-critical flow) with a value of $\chi$ below the instability threshold. 

The damping timescale also depends somehow on the distance to the linear instability threshold (Fig. \ref{fig:amplitude_velo}, bottom panel). In a second set of simulations, performed with $\delta\rho/\rho=30\%$ and several values of $\chi_0$, we observe that only the cases with an initial value of $\chi$ above the instability threshold are able to maintain turbulent convective motions over at least 15 advection timescales. 
Our results show that a single excitation is not sufficient to trigger a self-sustained instability when the flow is linearly stable. This can only occur in situations where the dynamics is continuously fed by non-linear perturbations such as in a simulation performed with a Cartesian grid which would produce significant noise \citep{ott13} or when SASI reaches large amplitudes \citep{cardall15,summa16}. Moreover, our results show that the perturbative analysis of \citet{foglizzo06} holds in cases where the perturbed flow is already in the non-linear regime. In more realistic 3D simulations, the $\chi$ parameter seems to be a reliable criterion to assess whether convection ($\chi\gtrsim3$) or SASI ($\chi\lesssim3$) dominates the dynamics shortly after the spherical symmetry is broken \citep{hanke13, takiwaki14, couch14, abdikamalov15}. The asymptotic value of $\chi$ in the turbulent convective regime will be discussed in Sections \ref{sec:quanti} and \ref{sec:reso} where we investigate the impact of dimensionality and resolution.

The exploration of the parameter space ($\chi_0$, $\delta\rho/\rho$) enables us to distinguish three different regimes:
\begin{itemize}
 \item linear instability if $\chi_0>\chi_{\rm crit}$,
 \item linear stability if $\chi_0<\chi_{\rm crit}$ and $\delta\rho/\rho \lesssim 1\%$,
 \item transient convection if $\chi_0<\chi_{\rm crit}$ and $\delta\rho/\rho \gtrsim 1\%$.
\end{itemize}
In the latter case, convection is only transitory and cannot be self-sustained. The instability damping timescale increases with higher perturbation amplitudes and depends in a less clear way on the nearness to the instability threshold (Fig. \ref{fig:amplitude_velo}).
Nevertheless, the interaction between ascending bubbles and a shock wave might be strong enough to trigger convection in a linearly stable flow due to a non-linear cycle different from SASI. 
It cannot be excluded that the inclusion of a shock wave allows new ways to sustain convection, at least for values of $\chi$ slightly below the linear instability condition.

\section{Impact of the dimensionality}
\label{sec:quali}

\subsection{Early non-linear phase}
\label{subsec:bubbles}

\begin{figure}
\centering
	\includegraphics[width=0.9\columnwidth]{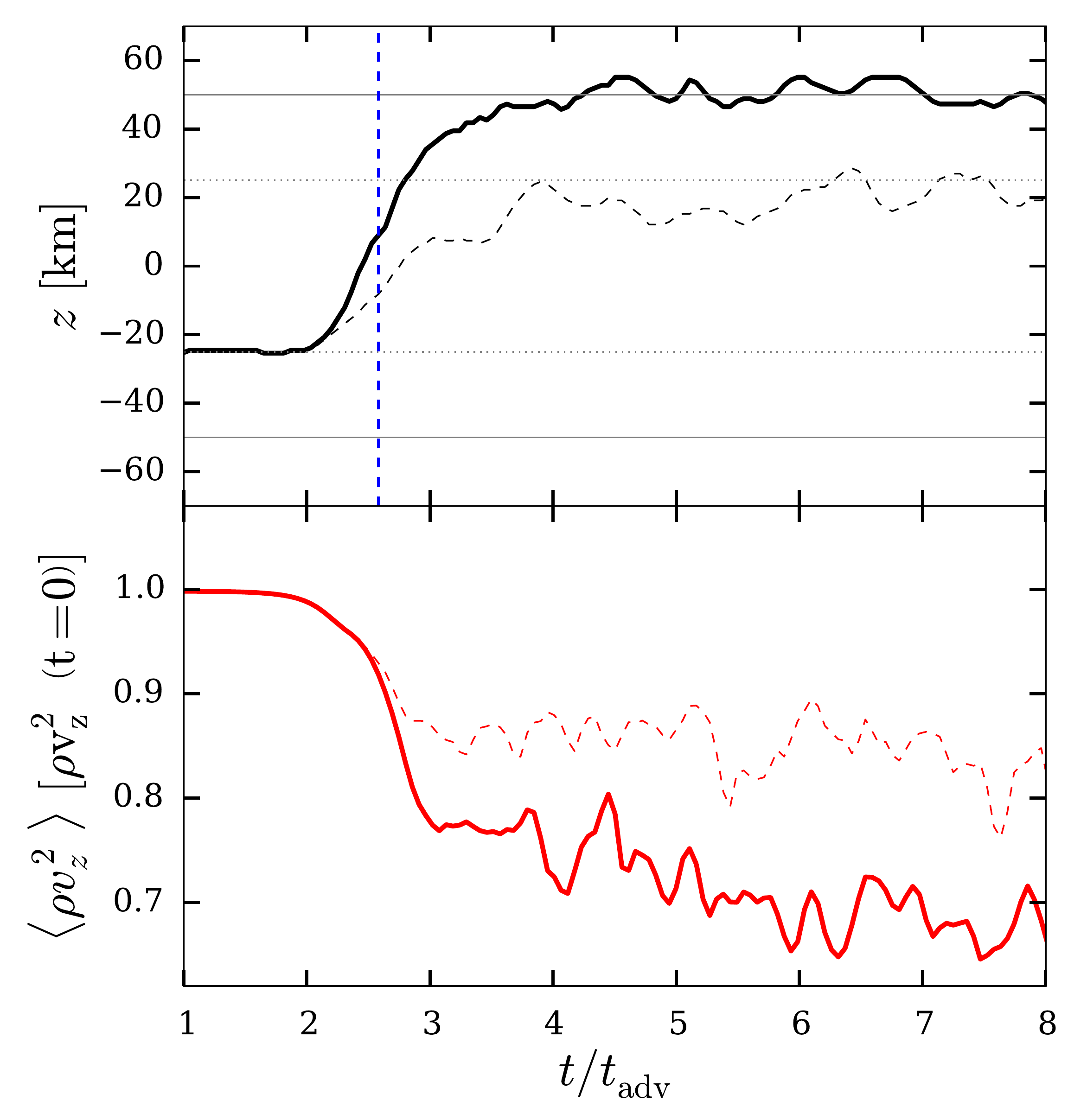}
    \caption{
    Altitude of the upper edge of the uppermost buoyant bubble (top panel) and ram pressure, defined as $\rho v_z^2$, (bottom panel) as a function of time for a model with $\chi_0=5$ and $\delta\rho/\rho=0.1\%$ simulated in 3D (thick curves) and in 2D (thin dashed curves). The horizontal black lines are defined in Fig. \ref{fig:perturbation}.
    The ram pressure is averaged between the planes $z=30\,\rm{km}$ and $z=45\,\rm{km}$. The vertical blue dashed line marks the time at which the ram pressure starts to deviate by more than 1\% between 2D and 3D.
    At this time, buoyant bubbles have already reached higher altitudes in 3D and this initial gap is not related to differences in terms ram pressure against the bubbles. 
    The later evolution is characterized by a lower ram pressure in 3D. This results from a more important acoustic feedback produced by convection which alters the subsonic upstream flow more strongly.
    }
    \label{fig:saturation}
\end{figure}

The first differences between 2D and 3D dynamics arise when perturbations reach non-linear amplitudes. If the instability is triggered by a low  amplitude perturbation in a flow where $\chi_0>\chicrit$, buoyant bubbles rise faster in 3D than in 2D (Fig. \ref{fig:snapshots}, top panels). In a simulation run with the parameters $\chi_0=5$ and $\delta\rho/\rho=0.1\%$, the highest 3D bubbles rise almost twice as fast against the flow and the instability saturates at a much larger amplitude (Fig. \ref{fig:saturation}). This is deduced from the time evolution of the location of the highest buoyant bubble. Its position is obtained by tracking the altitude of the uppermost negative radial entropy gradient which is due to the entropy contrast between the background flow (lower entropy) and the highest buoyant bubble (higher entropy). This transition is much sharper than the one related to the background flow.

A crude estimate of the velocity of the bubbles can be made by computing the terminal velocity $v_{\rm ter}$ which corresponds to the balance between the drag force ($S\rho_0 v^2_{\rm ter}$) and buoyancy ($V(\rho_0-\rho)g$):
\begin{equation}
 \label{eq:vlim}
 v^2_{\rm ter} = \frac{\left(\rho_0-\rho\right)V g}{\rho_0 S}.
\end{equation}
The terminal velocity increases with higher volume-to-surface ratio $V/S$. \citet{couch13a} interpreted earlier and easier explosions in their 2D simulations as a consequence of larger $V/S$ ratios compared to 3D. This explanation is based on different bubble shapes between 2D and 3D. Large-scale axisymmetric bubbles are considered to be more efficient to overcome ram pressure than 3D bubbles that fragment to small scales. However the ratio $V/S$ depends mostly on the vertical extension of the bubbles. 
For a bubble moving in the radial direction, the relevant surface is in the angular directions. The ratio $V/S$ is therefore approximately the vertical extension of the bubbles, within a factor of order unity depending on the exact shape.
For example, comparing a spherical bubble of a given radius with a toroidal one of the same radius, one finds that the terminal velocity is only slightly higher in 2D, by a factor of order unity $\sqrt{3\pi/8}$. Interpreting the role of dimensionality on CCSN simulations as a consequence of specific bubble properties is in conflict with our model where bubbles are clearly faster and reach higher altitudes in 3D than in 2D.

The acoustic feedback produced by the convective motions is able to alter the subsonic upstream flow. As a consequence, the ram pressure against the buoyant bubbles decreases with time (Fig. \ref{fig:saturation}). The upstream acoustic feedback is stronger in 3D, reducing the ram pressure more than in 2D simulations. However, we observe that when the ram pressure deviates by more than 1\% between 2D and 3D, the buoyant bubbles have already reached a larger altitude in 3D. This shows that the alteration of the upstream flow by convection motions cannot explain the early discrepancies in terms of altitudes reached by buoyant bubbles. It is conceivable that the acoustic feedback plays a role at a later stage to set the saturation amplitude of the instability. In a model with a shock wave and a supersonic upstream flow, a strong acoustic feedback would increase the post-shock pressure and induce a shock expansion. Our study suggests that this effect could be larger in 3D than in 2D.

Besides, our results seem consistent with conclusions drawn from studies focusing on the role of dimensionality in the Rayleigh-Taylor instability. The mixing zone is found to broaden faster in 3D as plume-like structures penetrate much deeper than 2D planar structures in simulations of incompressible flows without advection \citep{young01,anuchina04}. Similar conclusions were obtained in the context of the Rayleigh-Taylor mixing in the stellar envelope during the explosion of massive stars \citep{kane00, hammer10}. The growth of the Rayleigh-Taylor fingers is faster and the velocities of clumps of heavy elements are higher in 3D. Considering a more elaborate model of the competition between drag force and buoyancy, \citet{hammer10} showed that the temporal evolution of the $V/S$ ratio favours a faster rise of the bubbles in 3D. Our model underlines that during the early non-linear phase of the instability, the rise of buoyant bubbles and the widening of the mixing zone are faster in 3D, even when advection is considered.

\subsection{Heating efficiency}
\label{subsec:mixing}

\begin{figure}
\centering
	\includegraphics[width=\columnwidth]{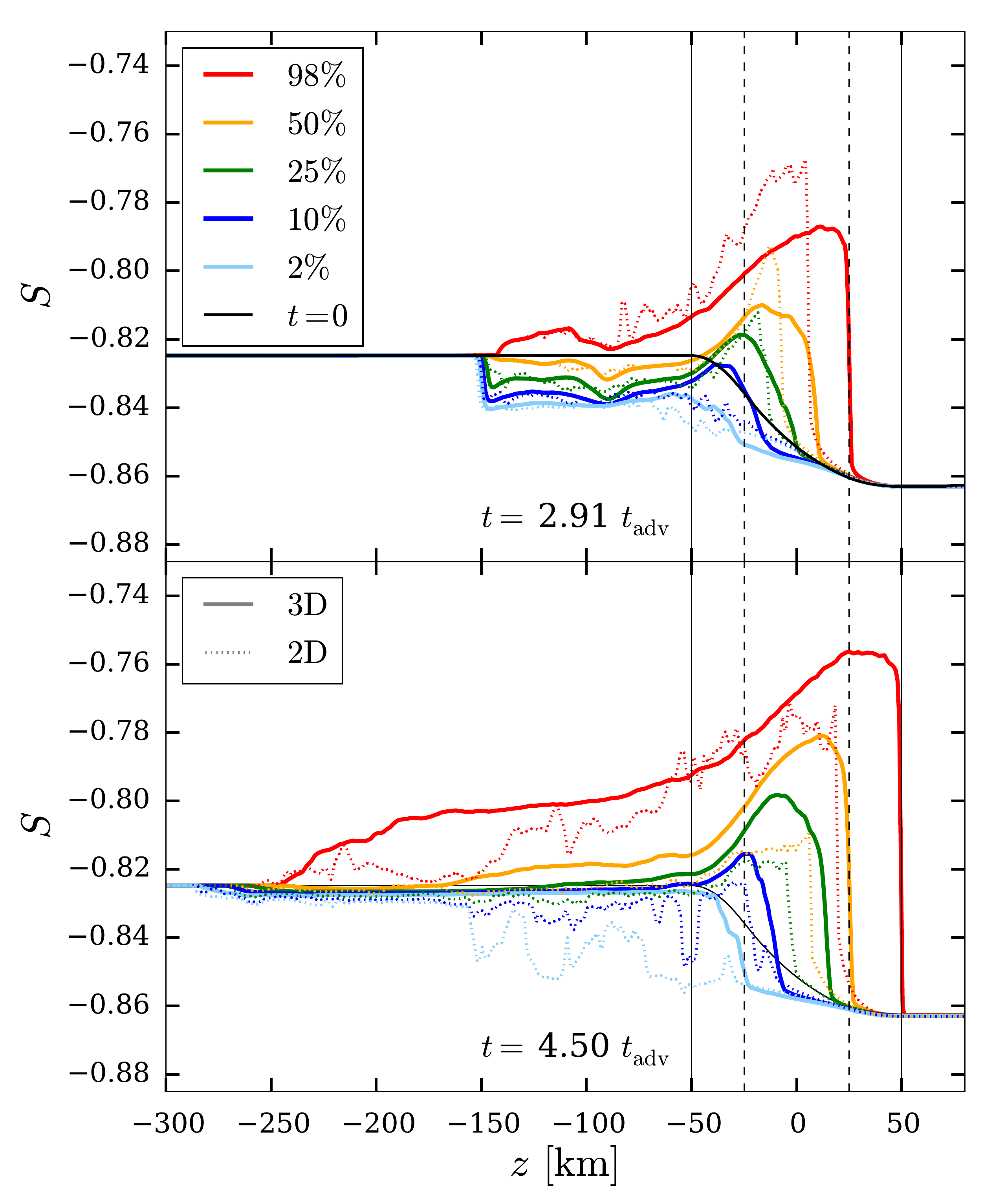}
    \caption{Vertical profiles of entropy obtained from 3D (thick curves) and 2D (thin dotted curves) simulations of a model with $\chi_0=5$ and $\delta\rho/\rho=0.1\%$, after 2.91 (top panel) and 4.5 (bottom panel) advection timescales. Each profile shows the entropy such that a given fraction of the numerical cells at a given altitude $z$ have a lower value. The entropy of the stationary flow is shown in black.}
    \label{fig:Sprofiles}
\end{figure}

Figure \ref{fig:Sprofiles} shows the time evolution of the spatial distribution of entropy in 2D and 3D. Each vertical entropy profile is defined such that a given fraction of the cells have a lower entropy value. In the early non-linear phase, the convective instability exhibits a wider mixing region in 3D while mixing is more localized in 2D. As a consequence, the vertical profiles related to high entropy values show more pronounced peaks in 2D. The rise of the 3D bubbles stops when they reach the upper edge of gain region above which gravity is switched off (Fig. \ref{fig:saturation}, top panel). After several advection times, the vertical profiles of entropy become higher in the whole gain layer in 3D (Fig. \ref{fig:Sprofiles}, bottom panel).

\begin{figure}
\centering
	\includegraphics[width=\columnwidth]{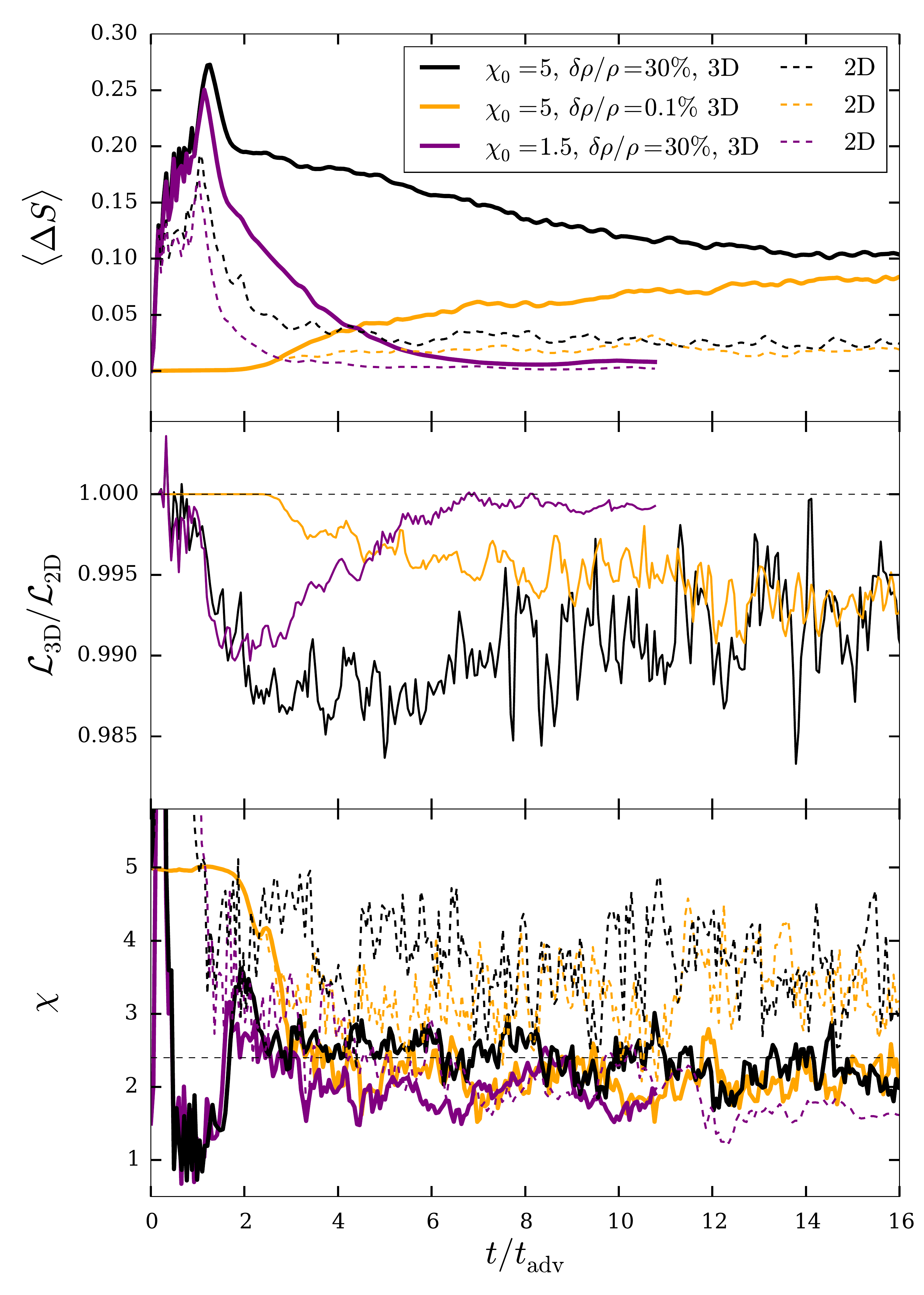}
    \caption{\textit{Top panel: } Time evolution of the average entropy variations in the gain layer in 2D (dashed curves) and in 3D (solid curves) for three sets of parameters. The variations are computed as: $\Delta S\equiv \langle S\rangle - \langle S\rangle_0$ where $\langle S\rangle$ is the average entropy at a given time and $\langle S\rangle_0$ at the initial time. 
    \textit{Middle panel: } Time evolution of the ratio of heating rates between 3D and 2D simulations. In all cases, the heating rate is slightly higher in 2D than in 3D.
    \textit{Lower panel: } Time evolution of the $\chi$ parameter in 2D (dashed curves) and in 3D (thick curves). In the regime of transient convection, the flow returns below the critical value $\chi\sim2.4$. In the linear instability regime, only the 3D dynamics is able to bring the flow to a sub-critical state.
    }
    \label{fig:timeevol}
\end{figure}

The generation of entropy in the gain region appears to be higher in 3D than in 2D (Fig. \ref{fig:timeevol}, top panel). At the end of the simulations performed with $\chi_0=5$, the entropy has increased about 4 times more in 3D than in 2D. We note that the amplitude of the initial perturbation does not affect the asymptotic dynamics because final entropy values are independent of the perturbation both in 2D and in 3D. In the regime of transient convection, the entropy increase is related to the amplitude of the perturbation and is higher in 3D (Fig. \ref{fig:timeevol}, top panel, purple curves). The damping timescale of the instability is much longer in 3D suggesting that perturbations may have a more profound impact on convection-dominated cases simulated in 3D than in 2D. A direct comparison of the heating rates, proportional to the mass enclosed in the gain region, does not explain the discrepancies in terms of entropy. It turns out that the rates are slightly higher in 2D throughout the simulations (Fig. \ref{fig:timeevol}, middle panel). The strong impact of dimensionality on the entropy variations is suggestive of a heating process that differs significantly between 2D and 3D. We defer the investigation on the origin of this additional heating to Section \ref{sec:quanti} where the turbulence induced by the instability is analysed in detail.

The morphology of the flow is strongly impacted by the dimensionality. In 2D, the flow is stirred by large vortices whereas 3D models undergo stronger turbulent mixing to small scales. Phase mixing is essential to disrupt large scale entropy perturbations even in the absence of heating (case $\chi_0 = 0$ of Fig. \ref{fig:amplitude_velo}, bottom panel). The dynamics generates large vortices in 2D whose vertical extension can reach the height of the gain layer. This can be seen as a consequence of a greater conversion of gravitational potential energy to kinetic energy by the Rayleigh-Taylor instability \citep{young01, cabot06}. In 2D, the interface between ascending motions and downflows is unperturbed. The downflows are able to sustain throughout the gain layer and channel lower entropy material located in the upper part of the gain layer to the downstream region where heating is absent (Fig. \ref{fig:snapshots}, bottom left). On the contrary, 3D downflows are disrupted by turbulent mixing before leaving the gain layer. Such a mixing is known to be more efficient in 3D in the case of the Rayleigh-Taylor instability \citep{cabot06, hammer10}. As the flow leaves the gain layer ($z=-50$ km), we observe that in 3D a smaller fraction of the numerical cells have an entropy lower than the stationary flow compared to 2D (Fig. \ref{fig:Sprofiles}, bottom panel). This implies that downflows are more efficient to channel cold entropy material outside of the gain region in 2D, while their fragmentation to small scales in 3D keeps buoyant material for longer times and enhances heating.

\begin{figure}
\centering
	\includegraphics[width=\columnwidth]{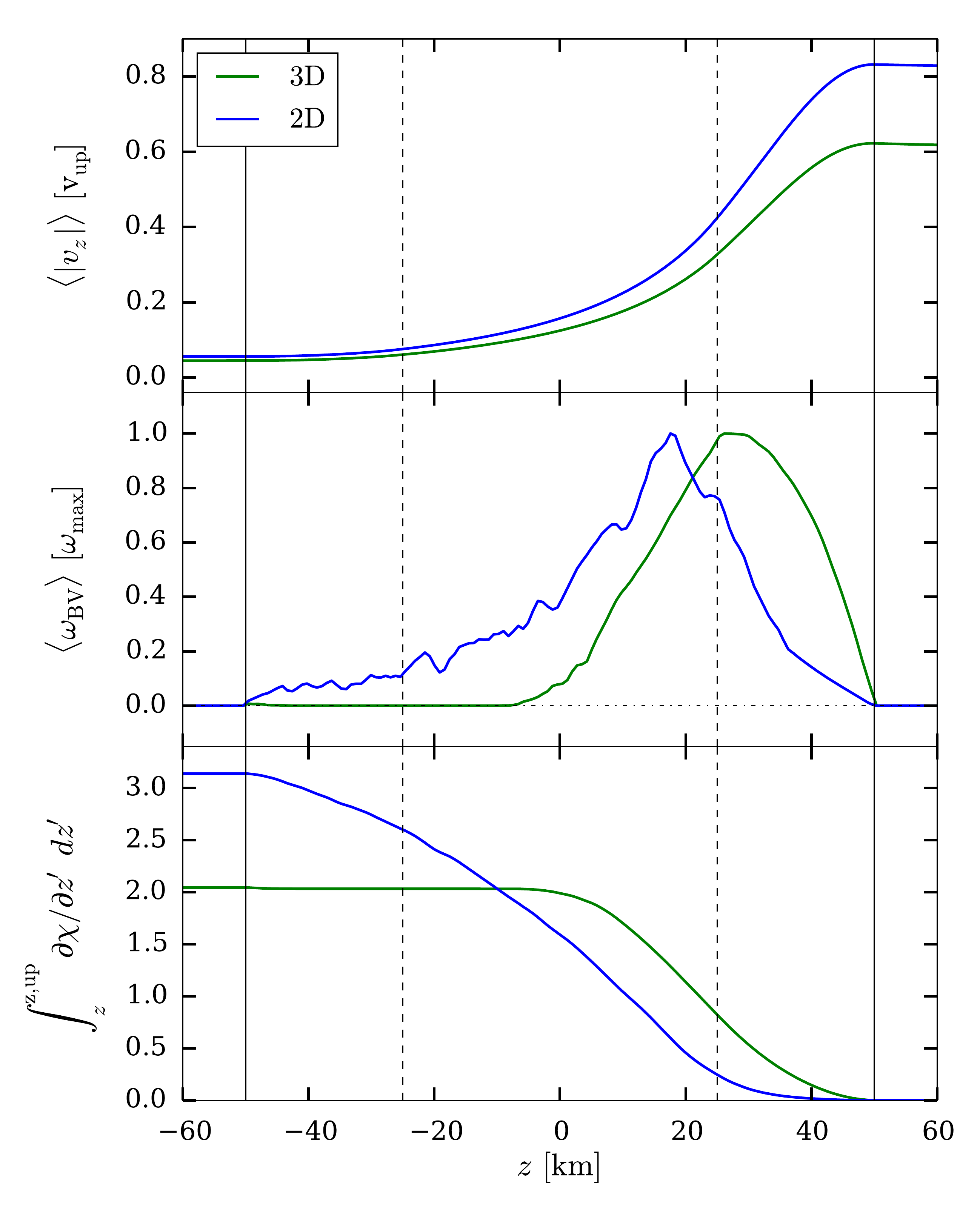}
    \caption{\textit{Top panel: } Vertical profiles of vertical velocity in the gain region of simulations performed in 2D (blue) and in 3D (green) with the parameters $\chi_0=5$ and $\delta\rho/\rho=0.1\%$. The quantities are averaged horizontally and over a period of 4 advection timescales. The vertical profiles are normalized by the upstream velocity of the stationary flow.
    \textit{Middle panel: } Vertical profiles of the Brunt-V\"{a}is\"{a}l\"{a} frequency.  The profiles are normalized to the maximum value of the Brunt-V\"{a}is\"{a}l\"{a} frequency in the gain region, noted $\omega_{\rm max}$. The upper part of the gain region is unstable to convection in 3D while the whole layer remains unstable in 2D.
    \textit{Bottom panel: } Vertical profiles of the cumulative integral of $\partial \chi/\partial z$ computed from the upper edge of the gain region, noted z,up. The higher $\chi$ values obtained in 2D seem to result from the contribution of the lower part of the gain layer.
    }
    \label{fig:profiles}
\end{figure}

The combine action of more efficient downflow braking and stronger upstream acoustic feedback in 3D produces smaller average downflow velocities (Fig. \ref{fig:profiles}). The advection time is thus slightly larger in 3D and this should in principle increase the heating efficiency compared to 2D. However, this is contradicted by the previous comparison of the heating rates.

\subsection{Asymptotic regime}
\label{subsec:chifinal}

The evolution of the $\chi$ parameter is related to the way the flow adjusts itself compared to the marginal stability limit (Fig. \ref{fig:timeevol}, bottom panel). The $\chi$ parameter is computed from horizontally averaged flows (Eq. \ref{eq:chi}) as proposed by \citet{fernandez14}.

When convection is triggered non-linearly by a strong perturbation ($\chi_0=1.5$, $\delta\rho/\rho=30\%$), the parameter initially takes large values during the advection of the perturbation. This is due to the strong entropy gradients contained in the perturbation (Fig. \ref{fig:perturbation}) which induce a larger value of $\chi$ for a brief period of time. Then, the value of $\chi$ drops much faster in 3D than in 2D. 
About $\sim 0.1\,\tadv$ after the perturbation has started to accrete through the gain layer, the 3D flow already reaches sub-critical values ($\chi\lesssim2.4$) because the perturbation is efficiently disrupted by strong turbulent mixing. It takes more than two advection times to reach the same level in 2D. In both cases the flow eventually returns to a sub-critical state. Besides, this comparison confirms that even in 3D a strong perturbation cannot lead to self-sustained convection when the flow is linearly stabilized (see Sect. \ref{subsec:sustained}).

The discrepancies are more remarkable in the linear instability regime (Fig. \ref{fig:timeevol}, bottom panel, black and orange curves). In  2D, the flow maintains a $\chi$ value well above the stability threshold ($\chi \gtrsim 3$) whereas in 3D, the flow adjusts itself to a slightly sub-critical flow ($\chi\approx2.2$). Slightly lower average vertical velocities in 3D (Fig. \ref{fig:profiles}, top panel) should in principle result in a higher value of $\chi$ than in 2D. However, turbulent mixing seems to be part of the explanation of the gap between 2D and 3D values of $\chi$. After the instability fully develops, a smaller fraction of the gain layer remains unstable to convection in 3D (Fig. \ref{fig:profiles}, middle panel). In 2D, the flow is dominated by large vortices which create a less stable configuration. The contribution of the lower part of the gain layer ($z\lesssim0$ km) seems responsible for the greater values of $\chi$ in 2D (Fig. \ref{fig:profiles}, bottom panel). These results suggest that any analytical description of turbulence induced by convection in CCSNe (e.g. \citealt{murphy11} in 2D) should reflect that difference when applied to our model.

\section{The role of turbulence}
\label{sec:quanti}

\subsection{Kinetic energy}
\label{subsec:kinetic}

Convection may facilitate the shock revival in several ways compared to 1D cases.
Non-radial motions induced by instabilities in multidimensional simulations increase the advection timescale and bring it closer to the time needed for a sufficient energy deposition in the gain layer \citep{murphy08}. Turbulent pressure generated by convective motions represents an additional pressure support which pushes the shock wave outward and lowers the critical luminosity required to achieve shock revival \citep{burrows95, murphy13, couch15, mueller15a}. Both effects may combine to enhance neutrino heating and favour multidimensional explosions. Comparing 2D and 3D simulations, \citet{couch15} concluded that the strength of turbulence is overpredicted in axisymmetric simulations.

\begin{figure}
\centering
	\includegraphics[width=0.95\columnwidth]{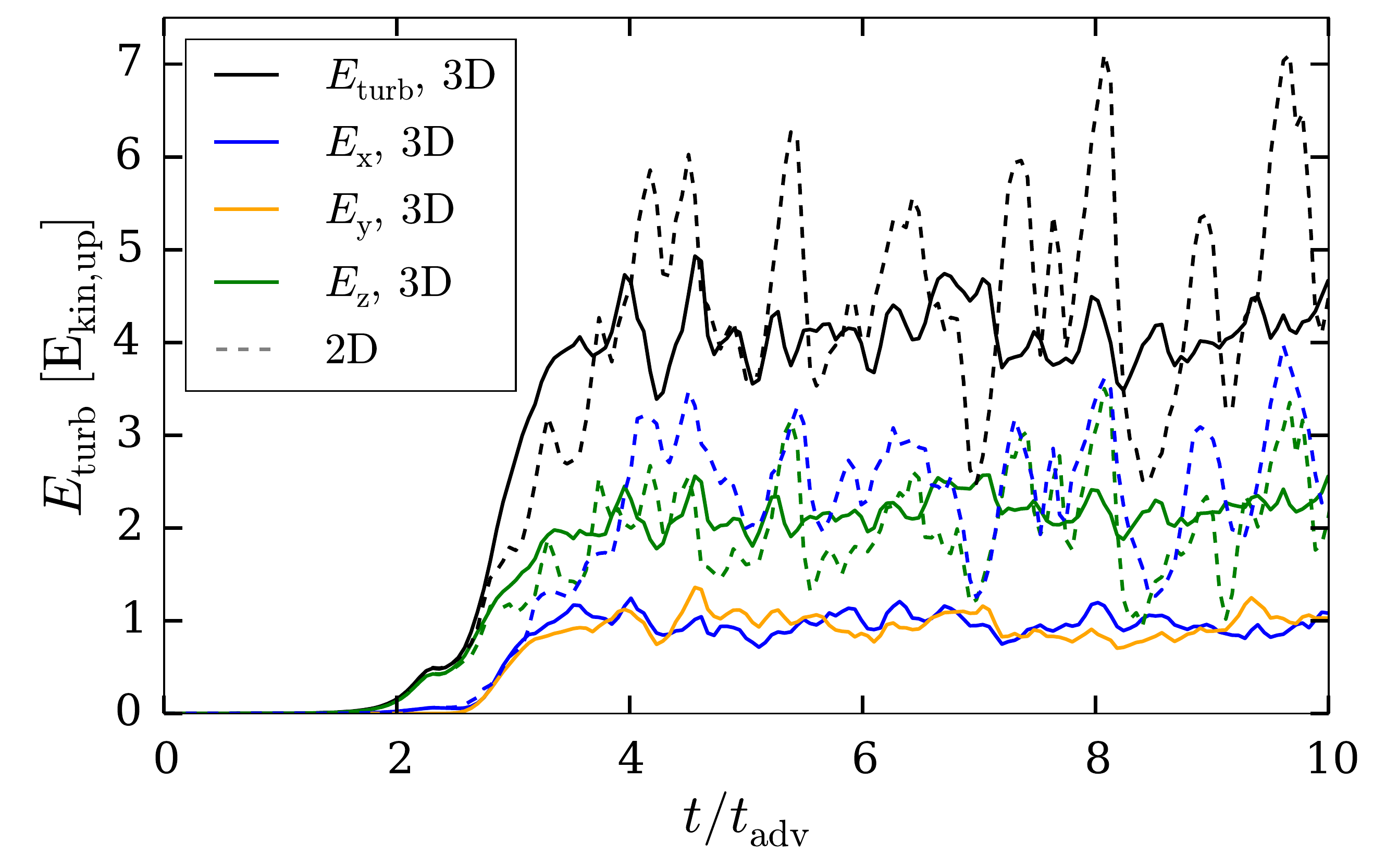}
    \caption{Turbulent kinetic energy integrated over the gain region as a function of time for simulations performed in 2D (dashed) and 3D (solid) with the parameters $\chi_0=5$ and $\delta\rho/\rho=0.1\%$.     
    The quantities are normalized by the kinetic energy of the stationary flow upstream from the gain layer, noted $E_{\rm{kin,up}}$.
    }
    \label{fig:ekin}
\end{figure}

The turbulent kinetic energy in the gain layer can be defined as:
\begin{equation}
 \label{eq:Eturb}
    E_{\rm turb} = \frac{1}{2}\rho\left[\left(v_z-\langle v_z\rangle\right)^2+v_x^2+v_y^2\right],
\end{equation}
where $\langle v_z\rangle$ corresponds to the horizontally-averaged vertical velocity. In our model, we find that the turbulent kinetic energy is larger in 2D by about $15\%$ during most of the simulation (Fig. \ref{fig:ekin}).  Besides, the turbulent kinetic energy is roughly in equipartition in 2D between the two directions. This can be seen as a consequence of the domination of large scale vortices. In 3D, we find that the horizontal contributions are in equipartition ($E_{x} \approx E_y$) and their sum amounts to the vertical contribution ($E_z \approx E_x+E_y$). This is the case in 3D models where buoyant motions are induced along a preferred direction \citep{murphy13}. Nevertheless, higher Reynolds stresses observed in 2D do not seem to support larger entropy values in our 3D simulations (Fig. \ref{fig:timeevol}).

The impact of dimensionality can also be investigated by comparing the turbulent energy cascades in 2D and 3D. To connect our study with previous ones \citep{hanke12, dolence13, couch14, handy14, couch15, abdikamalov15, radice15, radice16}, we compute the turbulent kinetic energy density spectra considering only horizontal motions. The kinetic energy density is decomposed into Fourier coefficients:
\begin{equation}
 \label{eq:fourier}
 \hat{E}_{\perp}\left(k_x,k_y\right) = \int_{V_g} \exp{\left({-2\pi i\frac{k_xx+k_yy}{L}}\right)}\sqrt{\rho\left(x,y\right)}v_{\perp}\left(x,y\right)dV,
\end{equation}
where $L$ represents the horizontal extent of the gain region ($L=300\,\rm{km}$ in our simulations), $V_g$ the volume of the gain region in which the source terms are at their full intensity ($\Psi(z)=1$) and $v_{\perp}$ the horizontal velocity component. The Fourier coefficients are computed at each discrete vertical coordinate of the grid and averaged over the volume $V_g$. Note that $k_y=0$ in 2D. The total horizontal kinetic energy density is then
\begin{equation}
 \label{eq:Ekk}
 E\left(k\right) = \sum_{k-1 < \|\left(k_x,k_y\right)\| \leq k} |\hat{E}_{\perp}\left(k_x,k_y\right)|^2.
\end{equation}
The Fourier spectra are normalized by $\sum_k E(k)$ to obtain unity integrals.

\begin{figure}
\centering
	\includegraphics[width=\columnwidth]{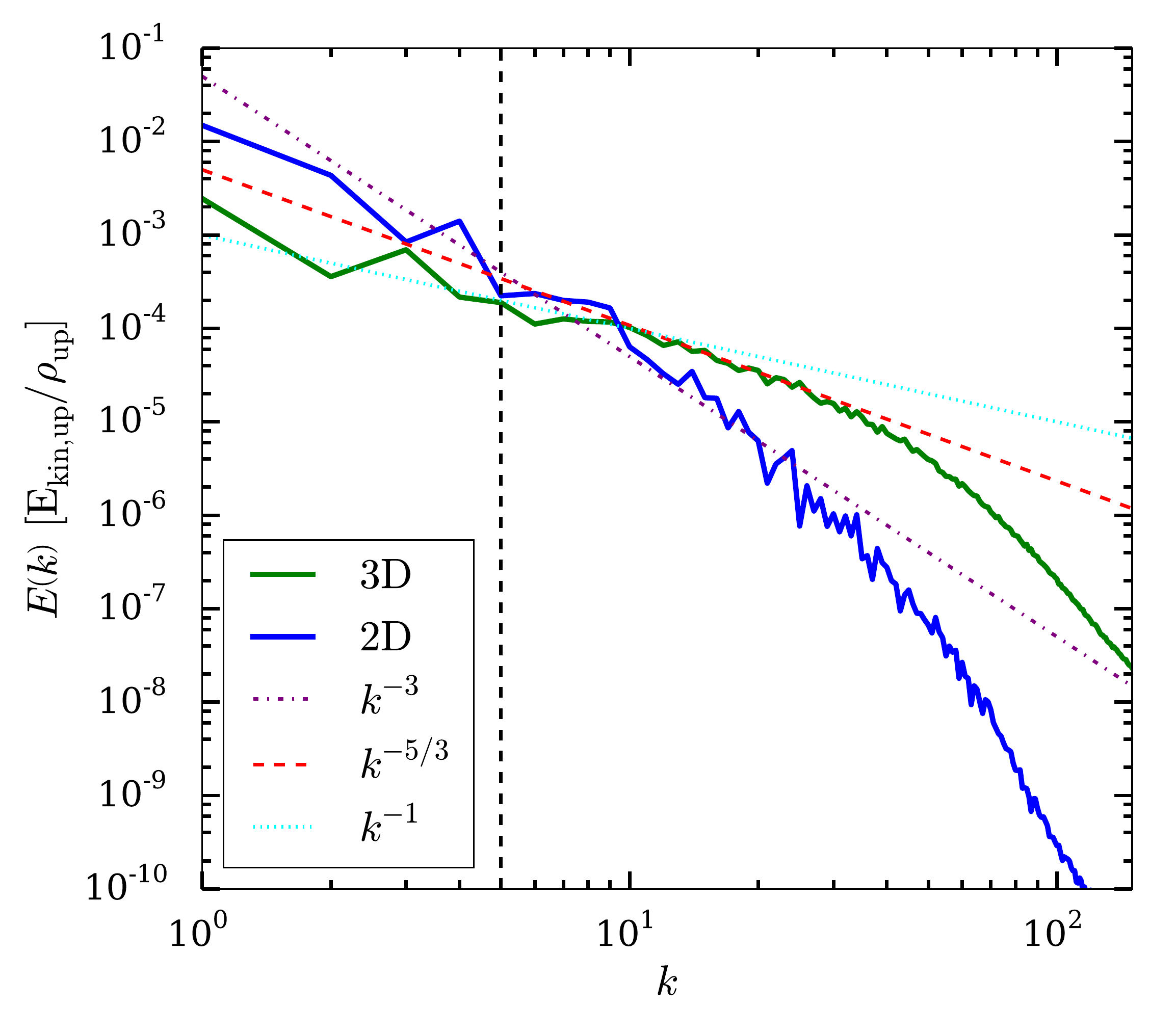}
    \caption{Spectra of turbulent horizontal kinetic energy density as a function of the wavenumber modulus $k$. The spectra are computed from 2D (blue) and 3D (green) simulations performed with the parameters $\chi_0=5$ and $\delta\rho/\rho=0.1\%$.
    The vertical black line represents the injection scale $k=5$ related to the initial perturbation in 2D and in 3D.
    In 2D, the reverse turbulent cascade follows a power-law $E(k)\varpropto k^{-3}$ (purple dot-dashed line) while the decay law in 3D follows $E(k)\varpropto k^{-1}$ at large spatial scales and  $E(k)\varpropto k^{-5/3}$ at intermediate ones.
    }
    \label{fig:spectrum}
\end{figure}

In 2D, the reverse turbulent cascade transfers kinetic energy to the largest spatial scales and this could explain why explosions are easier \citep{hanke12, couch14}. In the inertial range, enstrophy is transferred from large scales to the dissipation regime following $E(k) \varpropto k^{-3}$ (see \citet{kraichnan67, hanke12} for details). Such a trend seems to hold in our 2D simulations for $5\lesssim k \lesssim 30$ (Fig. \ref{fig:spectrum}), even though the identification of the inertial range is not obvious. In 3D, turbulence proceeds differently because a forward cascade transfers kinetic energy from large scales to the dissipation regime, following a decay power-law $E(k)\varpropto k^{-5/3}$ \citep{landau59}. Our simulations seem to show such a power-law decay (Fig. \ref{fig:spectrum}).
There is no consensus on the exact shape of the power-law in CCSN simulations. Some studies reported a decay law close to $E(k)\varpropto k^{-5/3}$ \citep{hanke12, handy14, radice16} while others observed a shallower decay $E(k)\varpropto k^{-1}$ \citep{dolence13, couch14, abdikamalov15}. 
Differences may result from a misidentification of the inertial range. A transition from a $k^{-1}$ scaling for $k \lesssim 15$ to a $k^{-5/3}$ scaling for $10\lesssim k \lesssim 40$ appears in our simulations.
The nature of the power-law in the inertial range reflects the way dissipation proceeds.
A limited numerical resolution could affect the shape of the power-law decay and only the finest resolutions, unachievable for current state-of-the-art CCSN simulations, could lead to a behaviour predicted by Kolmogorov's theory \citep{radice15,radice16}.
It can also be argued that the assumptions made in the classical theory do not apply to post-shock turbulence in CCSNe \citep{abdikamalov15}. 

\begin{figure}
\centering
	\includegraphics[width=0.85\columnwidth]{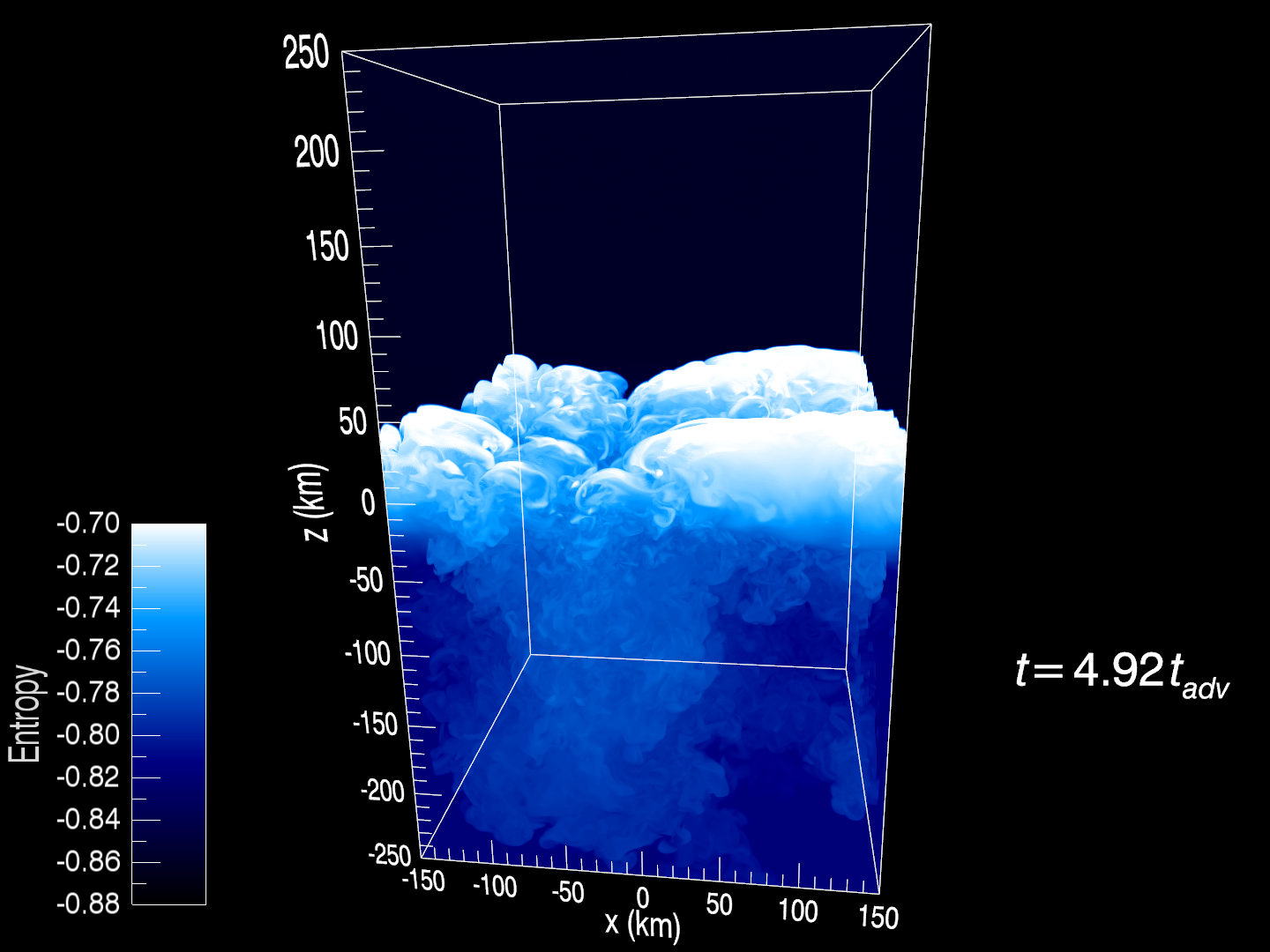}
    \caption{3D visualization of the entropy isosurfaces obtained from a simulation with $\chi_0=5$ and $\delta\rho/\rho=0.1\%$ after 4.92 advection timescales. Large and intermediate spatial scale structures can clearly be identified in the upper part of the gain layer. Animated versions of this figure are available in the online journal.
    }
    \label{fig:3Dvisu}
\end{figure}

Although the largest spatial scale structures are more pronounced in 2D, the spatial scales just below, between $k=3$ and $k=10$, seem almost equally favoured in our 2D and 3D simulations. This is consistent with the large entropy structures seen in the upper part of the gain layer (Fig. \ref{fig:snapshots}, bottom right). The presence of large scale structures can also be observed in Figure \ref{fig:3Dvisu} which provides a 3D visualization of entropy isosurfaces of the simulation performed with $\chi_0=5$ and $\delta\rho/\rho=0.1\%$, when the instability is fully developed. Nevertheless, the discrepancies in terms of turbulent kinetic energy do not seem to be directly connected with a stronger heating that occurs in our 3D simulations.

\subsection{Turbulent dissipation}
\label{subsec:dissipation}

In this section, we use a different approach to investigate the impact of dimensionality on turbulence. Following the work of \citet{murphy11, murphy13}, we employ a mean-field decomposition of the entropy equation into background and turbulent flows to identify the driving agent of the discrepancies in terms of heating. To this end, we decompose the variables into mean and fluctuation using two methods depending on the quantities. The Reynolds decomposition is defined according to $\phi = \overline{\phi}+\phi^{\prime}$ where the mean-field average of the fluctuation $\phi^{\prime}$ is $\overline{\phi^{\prime}}=0$. The Favrian decomposition corresponds to $\psi=\tilde{\psi}+\psi^{\prime\prime}$. Here the Favrian averaged quantity is a density-weighted average: $\tilde{\psi}=\overline{\rho\psi}/\overline{\rho}$. The Favrian fluctuation is such that $\tilde{\psi^{\prime\prime}}=0$. This second decomposition appears often to be more convenient for terms containing density in equations. We refer the reader to \citet{mocak14} for a complete overview of the derivation of hydrodynamics equations using the Reynolds-Averaged Navier Stokes method.

The entropy equation is written as:
\begin{equation}
\label{eq:Sequa}
 \frac{\partial\left(\rho S\right)}{\partial t} = -\nabla\cdot\left(\rho S v_z\right) + \frac{\rho\dot{q}}{T} + \frac{\rho\epsilon}{T},
\end{equation}
where $\dot{q}$ is the local heating rate, $T$ the temperature and $\epsilon$ the heat due to turbulent dissipation of kinetic energy. 
Applying a mean-field decomposition to Eq. (\ref{eq:Sequa}), we obtain:
\begin{equation}
\label{eq:Sdecomp}
 \frac{\partial \bar{\rho}\tilde{S}}{\partial t} = -
 \nabla.\left(\bar{\rho}\tilde{v_z}\tilde{S}\right) - 
 \nabla.\left(\overline{\rho v^{''}_z S^{''}}\right) +
 \overline{\left(\frac{\rho \dot{q}}{T}\right)} +
 \overline{\left(\frac{\rho \epsilon}{T}\right)}.
\end{equation}
The derivation is detailed in appendix \ref{sec:appendixA}.
These terms are labelled from (i) to (v) and represent:
\begin{enumerate}
 \item $\partial \bar{\rho}\tilde{S}/\partial t$, the generation rate of entropy,
 \item $\nabla.\left(\bar{\rho}\tilde{v_z}\tilde{S}\right)$, the entropy flux due to the mean flow across the gain layer,
 \item $\nabla.\left(\overline{\rho v^{''}_z S^{''}}\right)$, the entropy flux due to the turbulent flow across the gain layer,
 \item $\overline{\left(\rho \dot{q}/T\right)}$, the heating rate due to neutrino absorption,
 \item $\overline{\left(\rho \epsilon/T\right)}$, the turbulent dissipation.
\end{enumerate}
This last term is evaluated from Eq. (\ref{eq:Sdecomp}) and can also be approximated by $W_{\rm b}/T$ \citep{murphy13} where $W_{\rm b}$ is the buoyancy work, defined as:
\begin{equation}
\label{eq:Wb}
 W_{\rm b} = \int \overline{\rho^{\prime}v_z^{\prime}}g dV.
\end{equation}
The term $W_{\rm b}/T$ is labelled as (vi) in the following.
It approximates the turbulent dissipation if all the energy injected into turbulence by the buoyancy force is dissipated into heat. This would be expected in a quasi-steady state situation if the turbulent energy advected out of the gain region can be neglected.

\begin{figure}
\centering
	\includegraphics[width=0.9\columnwidth]{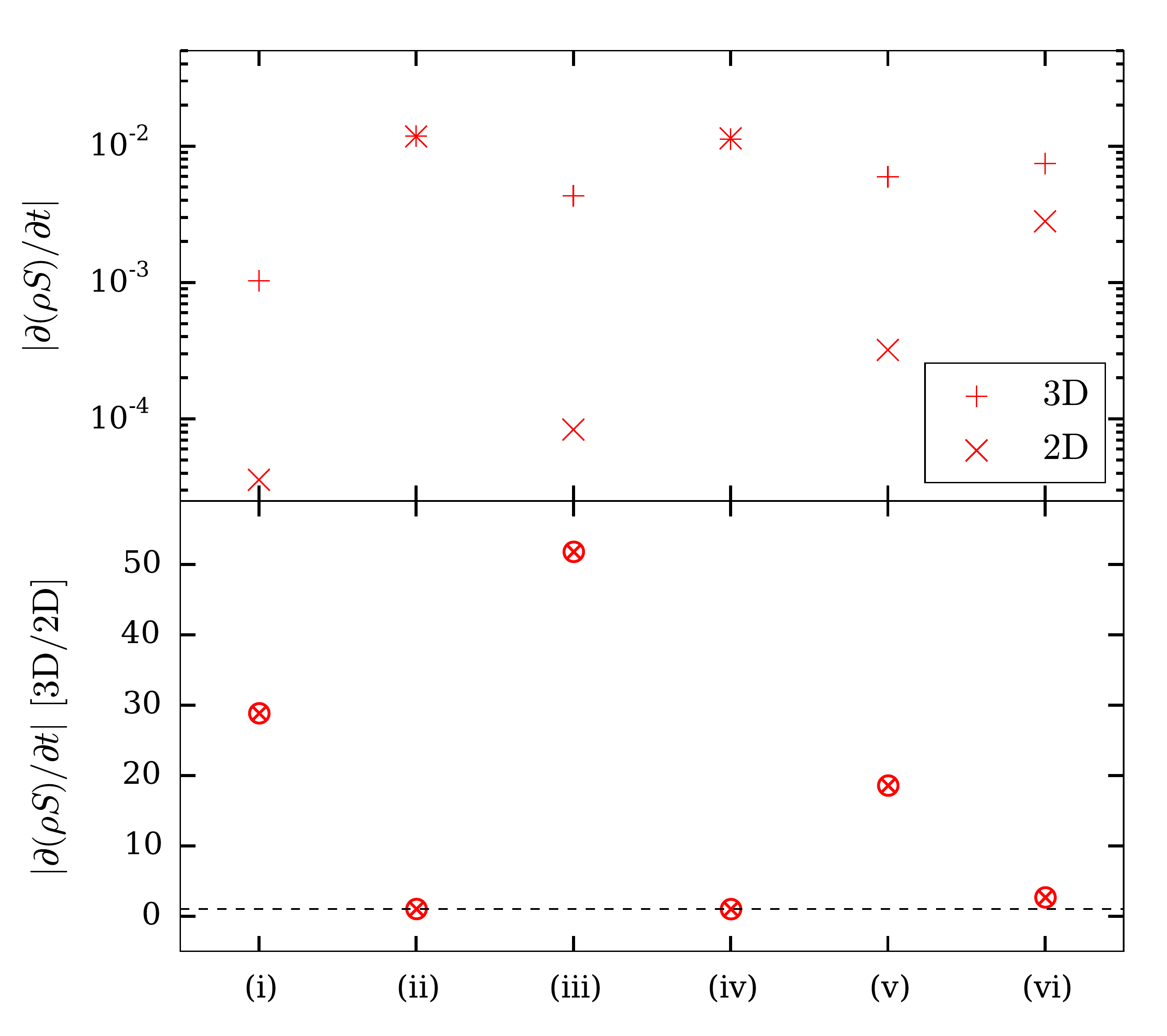}
    \caption{Time averaged values of entropy production terms in a case with $\chi_0=5$ and $\delta\rho/\rho=0.1\%$. The definitions of the six terms are given in the text.
    \textit{Top panel:} The values obtained in the 3D run (symbols +) are compared to the ones from the 2D run (symbols x). All terms are positive except (iii) in 2D.
    \textit{Bottom panel:} Ratios of 3D to 2D average values for the six terms. The horizontal dashed line denotes the ratio of one.
    }
    \label{fig:dissip}
\end{figure}

The terms (i) to (vi) are averaged horizontally and over four advection times during the fully non-linear regime. These terms are then integrated over the height of the gain region and compared to pinpoint which ones play a significant role in the entropy production (Fig. \ref{fig:dissip}). Note that all the terms are positive except the entropy flux due to the turbulent flow in 2D.
The neutrino heating rate appears to be the main contributor to the average entropy. It is almost balanced by the entropy flux due to the mean flow which accounts negatively to the entropy production (Eq. \ref{eq:Sdecomp}). These terms are almost independent of the dimensionality and in particular $\dot{q}$ is only slightly higher in 2D, as already shown in Fig. \ref{fig:timeevol}.

The second main heating source is the turbulent dissipation of kinetic energy (term (v)). In 3D, it accounts for almost half of the heating due to neutrinos. The entropy generation rate is around twenty times higher in 3D than in 2D (term (i)), so is the ratio of turbulent dissipation. This suggests that the additional entropy production in our 3D simulations results from the contribution of the turbulent dissipation. In 3D, the term related to the turbulent flux is slightly lower than the turbulent dissipation but accounts negatively to the average entropy. In 2D, this quantity only marginally contributes to increase the average entropy since it is the smallest term.

The buoyancy work is almost balanced by turbulent dissipation in 3D as shown by \citep{murphy13}. This is not satisfied in 2D, where the buoyancy work is an order of magnitude larger than the turbulent dissipation. The much lower dissipation in 2D may be interpreted as a consequence of the large scale vortices being expelled from the gain layer (Fig. \ref{fig:snapshots}) whereas those are dissipated in 3D. The latter is more efficient to dissipate kinetic energy because of the forward energy cascade to small scales. Turbulent dissipation is less likely in 2D since large vortices and downflows are not efficiently disrupted.

Using various approaches and setups, it was shown that turbulent motions generated by hydrodynamical instabilities in multi-dimensional simulations can account for a reduction of the critical neutrino luminosity of $20-30\%$ compared to 1D \citep{murphy08,hanke12,mueller15a,fernandez15}. The hierarchy between 2D and 3D is less clear and requires analysis of the properties of turbulence in the gain region in order to pinpoint which ingredients play a crucial role. Applying the turbulence model developed by \citet{murphy11,murphy13} on the explodability condition \citep{burrows93,murphy17}, \citet{mabanta18} conjectured that turbulent dissipation of kinetic energy is the dominant effect in the reduction of the critical luminosity in multidimensional simulations compared to 1D. Our results tend to support their hypothesis and the higher rate of entropy production in 3D seems to be connected to a more efficient dissipation of kinetic energy into heat. Our approach shows that such a physical phenomenon is less effective in 2D due to the particular tendency of favouring large spatial scales.

\section{Dependence on numerical resolution}
\label{sec:reso}

\begin{figure}
\centering
	\includegraphics[width=\columnwidth]{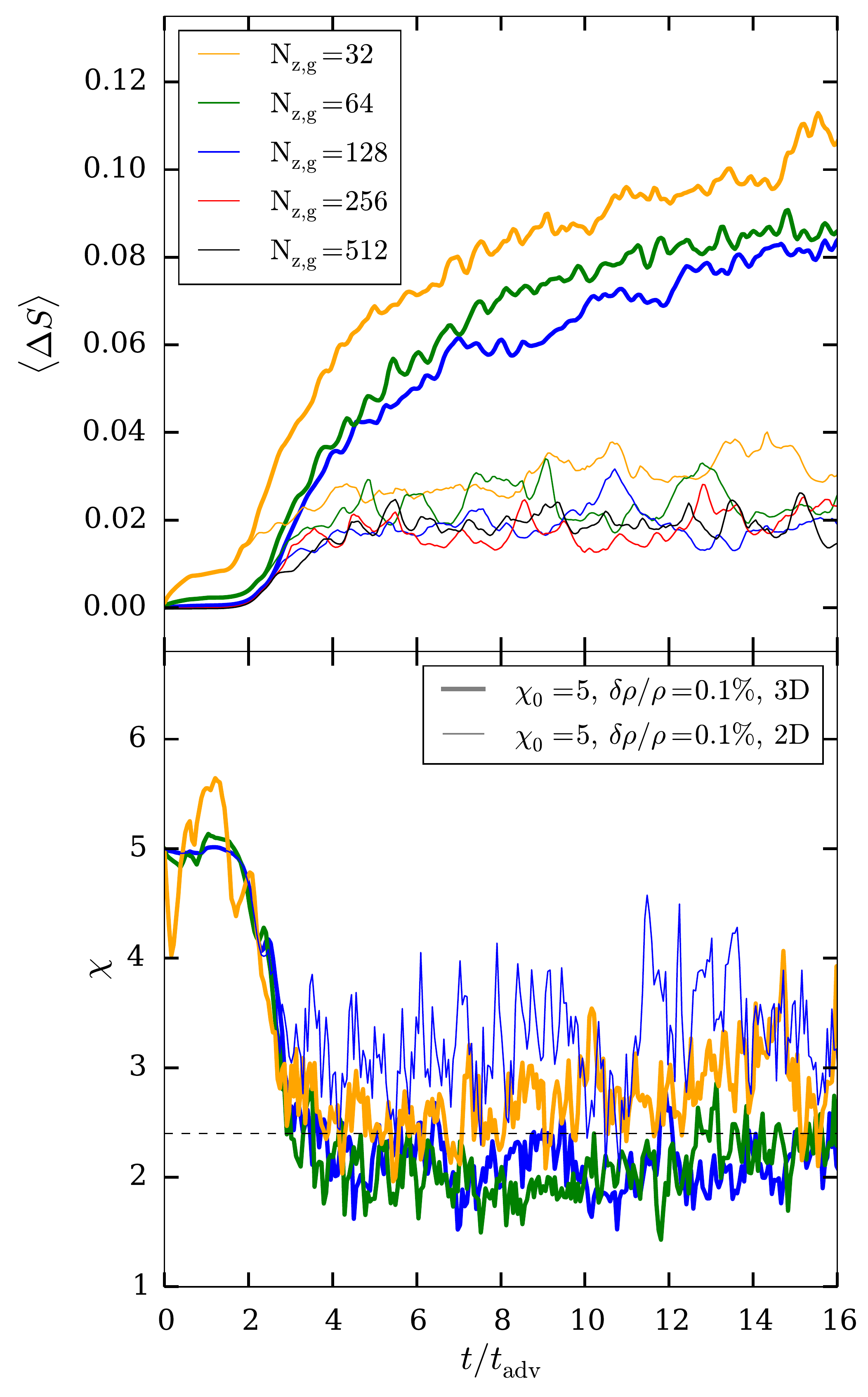}
    \caption{Results obtained from 2D (thin curves) and 3D (thick curves) simulations performed with the parameters $\chi_0=5$ and $\delta\rho/\rho=0.1\%$ for different resolutions labelled according to the number of vertical cells in the gain layer.
    \textit{Top panel: } Time evolution of the average entropy in the gain layer in 2D and 3D simulations.
    \textit{Lower panel: } Time evolution of the $\chi$ parameter in 2D and 3D simulations. The flow adjusts itself below the marginal stability (horizontal dashed line) in 3D for high enough resolutions.
    }
    \label{fig:time_reso1}
\end{figure}

\begin{figure}
\centering
	\includegraphics[width=\columnwidth]{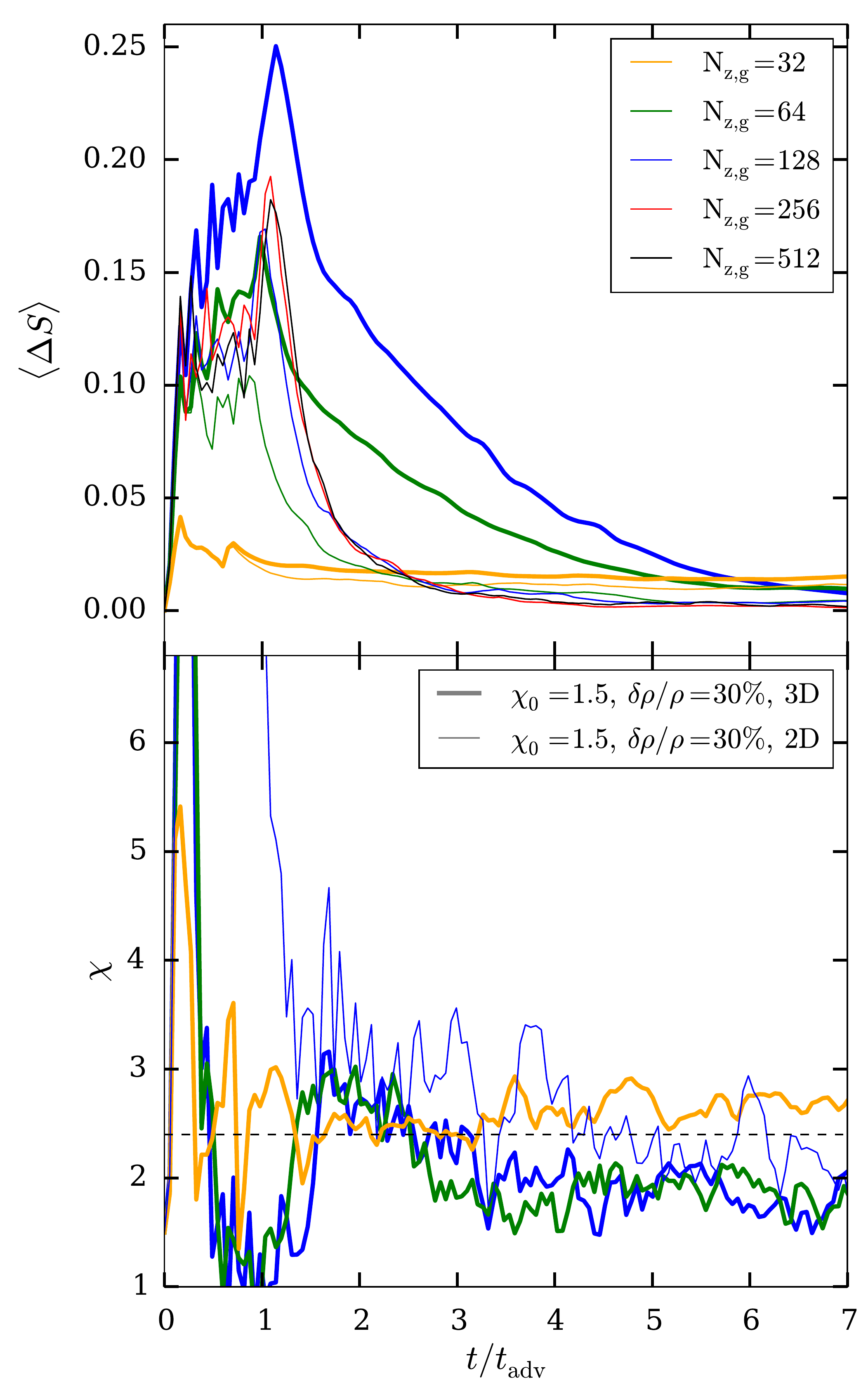}
    \caption{Same as in Figure \ref{fig:time_reso1} but for the parameters $\chi_0=1.5$ and $\delta\rho/\rho=30\%$.
    }
    \label{fig:time_reso2}
\end{figure}

We explore the impact of the numerical resolution on the 2D and 3D dynamics. The default resolution used in the previous sections is such that the uniform grid contains $\rm{N_{z,g}=128}$ vertical cells in the gain layer. The numerical resolutions considered in this section are $\rm{N_{z,g}=\{32,64,128,256,512\}}$ in 2D and $\rm{N_{z,g}=\{32,64,128\}}$ in 3D. The resolution is changed accordingly in the whole computational domain and in all directions to maintain the numerical cell size identical in each direction. For reference, our highest 3D resolution corresponds to twice the highest resolution considered in the studies of \citet{abdikamalov15, roberts16} and to the resolution labelled as ``2x'' in the work of \citet{radice16}. Figures \ref{fig:time_reso1} and \ref{fig:time_reso2} respectively give an overview of the influence of resolution on the dynamics in the linear instability regime and in the regime of transient convection. The main difference between the two regimes is that numerical resolution has a greater impact when the instability is triggered by a perturbation of a large amplitude.

Regarding the linear instability regime, the higher entropy production rate is not dramatically dependent on the resolution either in 2D or in 3D (Fig. \ref{fig:time_reso1}, top panel). On the one hand, if the number of vertical cells in the gain layer is too low, e.g. $\rm{N_{z,g}=32}$, the instability is triggered slightly earlier. This is probably due to the relaxation of the initial flow on the grid which deviates from the stationary solution. Besides an earlier onset of the instability at this low resolution, the entropy values reached may be artificially higher than for other resolutions. This suggests that the dynamics cannot be properly simulated in our model with such a resolution. On the other hand, if $\rm{N_{z,g}\geq64}$ the linear growth of the instability and the entropy production are very similar in all cases. This shows that the discrepancies between 2D and 3D concerning the entropy variations discussed in Section \ref{subsec:mixing} are not resolution-dependent. The lowest 3D resolution case produces a value of $\chi$ which is similar to 2D simulations (Fig. \ref{fig:time_reso1}, bottom panel) and the flow does not adjust itself to a slightly super-critical state. Other resolutions are not shown in 2D because their influence on $\chi$ is very minor. In 3D, a resolution such that $\rm{N_{z,g}\geq64}$ seems sufficient to bring the flow to a sub-critical state.

In Section \ref{subsec:mixing} we pointed out that 3D favours a larger entropy production when convection is triggered by a large amplitude perturbation. The same conclusion seems to hold with a higher resolution both in 2D and in 3D (Fig. \ref{fig:time_reso2}, top panel). The maximum entropy value reached rises with increasing dimensionality and resolution. Moreover, the damping timescale is also longer in 3D and with increasing resolution. In 2D, clear signs of convergence are witnessed between the three highest resolutions considered. It is unclear if a similar trend would be obtained in 3D. Only a too coarse resolution, e.g. $\rm{N_{z,g}=32}$, prevents the flow from reaching a sub-critical state that is expected for a simulation where $\chi_0=1.5$ (Fig. \ref{fig:time_reso2}, bottom panel). 

\begin{figure}
\centering
	\includegraphics[width=\columnwidth]{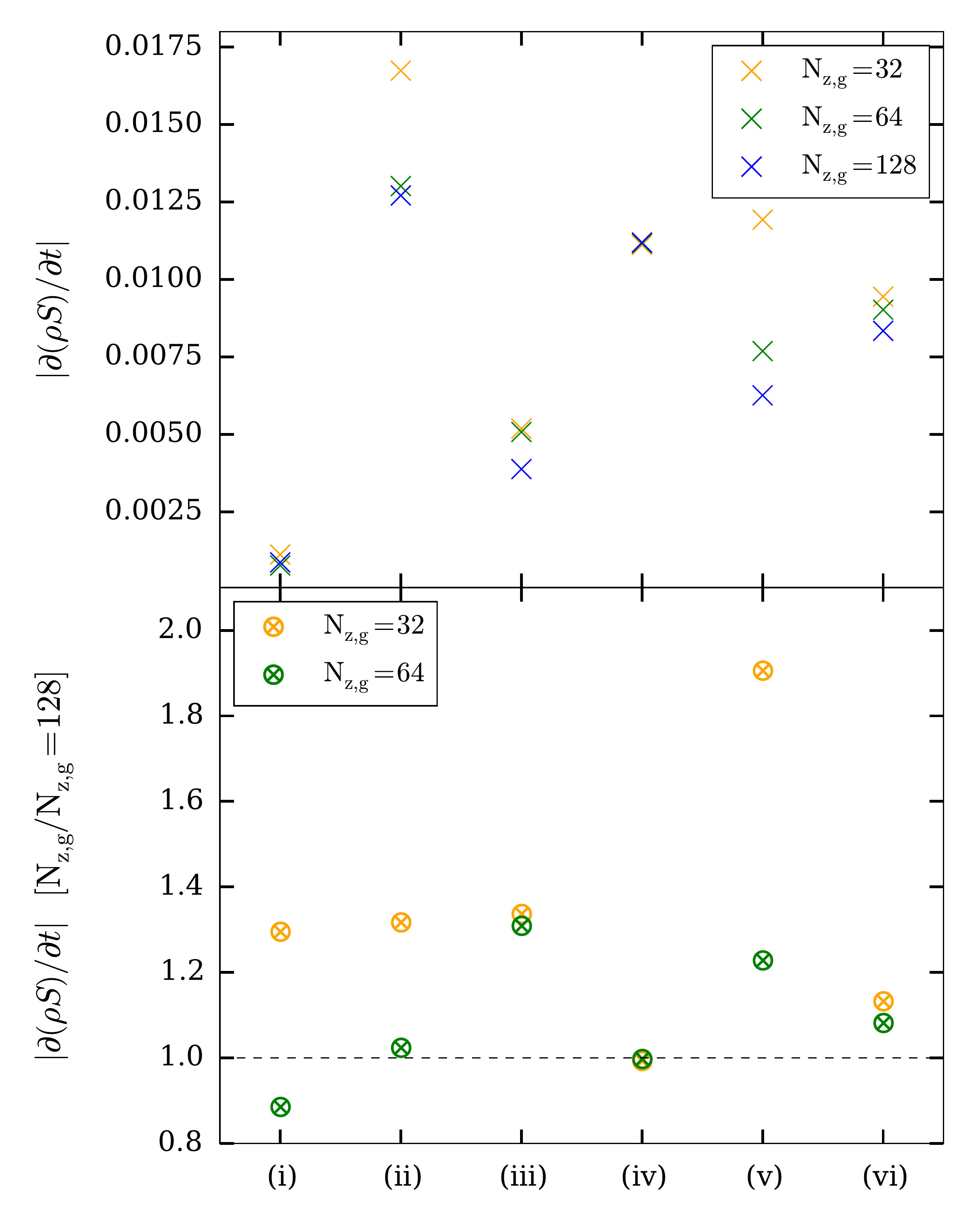}
    \caption{\textit{Top panel:} Time averaged values of entropy production terms in a case with $\chi_0=5$ and $\delta\rho/\rho=0.1\%$ run in 3D with different resolutions, labelled according to the number of vertical cells in the gain layer. The definitions of the six terms are given in section \ref{subsec:dissipation}. All the terms are positive.
    \textit{Bottom panel:} Same as above but normalized by the values obtained in the highest resolution run ($\rm{N_{z,g}=128}$). The horizontal dashed line denotes the ratio of one.
    }
    \label{fig:dissip_reso}
\end{figure}

We now apply the same mean-field decomposition of the entropy equation (\ref{eq:Sdecomp}) on 3D simulations of the linear instability regime with different numerical resolutions. The results indicate that except for turbulent dissipation at the lowest resolution, the different terms do not vary by more than $30-40\%$ (Fig. \ref{fig:dissip_reso}). This may be expected from the minor variations seen in entropy production rates with resolution (Fig. \ref{fig:time_reso1}, top panel). 
\citet{abdikamalov15} showed that even for their highest resolution of 66 radial cells in the gain layer, the inertial range of the turbulent energy cascade is barely resolved, thus favouring greater dissipation. This could explain why turbulent dissipation decreases with finer resolution in our simualtions.
Nevertheless, our study suggests that simulating turbulent dissipation with a level of $\sim20\%$ accuracy seems achievable with a resolution of about 100 vertical cells in the gain layer (see \citet{mueller16a} for a discussion). Even if a resolution ten times finer than ours could be needed to accurately resolve the inertial range, it appears that such an improvement does not affect global quantities such as the shock radius \citep{radice16}. The influence of resolution on the turbulent dissipation and the entropy variations in a convection-dominated case seems to be already captured by our set of 3D simulations with limited numerical resolutions.

\section{Discussion}
\label{sec:discussion}

Our comparative study of the 2D and 3D convective instability may be affected by some modelling ingredients used in this work. In the following, we discuss the differences with more realistic models and the implications of our results for self-consistent CCSN simulations. Our model consists of a static gain layer that is not delimited by a shock wave above and a stratified cooling layer below. The absence of a shock wave lowers the instability threshold and favours the instability for long wavelength perturbations (Fig. \ref{fig:kmax}). Replacing the constant inflow upper boundary condition by a shock wave would provide a continuous source of perturbations. The acoustic retroaction produced by the dynamics in the central region would deform the stationary shock and generate new entropy-vorticity perturbations that may seed the instability. Our study indicates that the upstream acoustic feedback is stronger in 3D and this may correspond to larger shock expansions than in 2D. In Section \ref{sec:regimes}, we considered an extreme case of a perturbation amplitude of 30\% to assess its impact on a linearly stabilized flow. Pre-shock density fluctuations related to combustion inhomogeneities may reach up to $10-15\%$ \citep{mueller15a}. The ones related to SASI may be lower with an upper estimate of $10\%$ according to the models of \citet{kazeroni16}. Grid induced perturbations could produce fluctuations of $1-10\%$ \citep{ott13}. The lower the amplitude, the shorter the damping timescale of the instability. According to Figure \ref{fig:amplitude_velo}, the damping timescale may be of the order of only a few advection times for realistic perturbation amplitudes. This implies that a continuous feeding by perturbations could lead to turbulent convection even if the flow is not linearly unstable to convection. This is the case when secondary convection is seeded by large amplitude SASI oscillations \citep{cardall15, summa16}. Such a situation can also occur when buoyant bubbles repeatedly push the shock or when numerical features produce noise continuously.

The absence of a cooling layer and a shock wave could impact the way convection modifies the flow. In their 2D models, \citet{fernandez14} obtained asymptotic values of $\chi$ below the instability threshold for non-exploding, convection-dominated cases whereas this value remains well above the threshold in our 2D simulations. The dynamics seems more efficient at stabilizing the flow in 2D in their study (\cite{fernandez14}, Fig. 7) than in our case (Fig. \ref{fig:profiles}). The 3D dynamics may be less impacted by the boundary conditions since the flow reaches a sub-critical state. (Fig. \ref{fig:time_reso1}). Our 2D results also differ from more realistic models regarding the balance between buoyancy work and turbulent dissipation. The equilibrium is almost reached in 3D but the buoyancy work is one order of magnitude larger than dissipation in 2D (Fig. \ref{fig:dissip}). Such a balance was reached in the 2D simulations of \citet{murphy11}. A possibility is that 2D vortices are not dissipated as much as in a case where they are kept inside the gain region due to the stabilizing cooling layer. Our setup may amplify the difference between 2D and 3D in terms of the amount of kinetic energy dissipated into heat.

Similarly, our setup may artificially widen the gap between 2D and 3D in terms of entropy production rates. \citet{hanke12} observed only slightly higher average entropy values in the gain layer of their 3D simulations compared to their 2D ones. The difference was more pronounced in the work of \citet{dolence13}, but of the order of 10\%. The latter study, using a Cartesian grid, might have been more prone to develop grid perturbations that could have artificially seeded convection. The work of \citet{hanke12} employed a spherical grid that generates less fluctuations. It is difficult to speculate whether a Cartesian grid would necessarily produce more heating via turbulent dissipation due to higher amplitude perturbations because the discrepancies between different groups could equally come from several specific approximations used in each study such as the neutrino transport or the microphysics. It should be noted that the average entropy value does not seem to represent a reliable indicator of the readiness of explosion \citep{hanke12, couch13a}.

Our comparative study of 2D and 3D neutrino-driven convection employs only Cartesian coordinates, neglecting the effects of geometric convergence, while previous works compared 2D axisymmetric geometry with 3D spherical or Cartesian geometries. \citet{foglizzo06} showed that the linear instability threshold is not affected by the geometry in their similar model including a shock wave. The same applies to the degree of the first unstable mode. However, conclusions cannot be drawn beyond the linear phase. Our study sheds light on a higher entropy creation in 3D, possibly related to more efficient turbulent dissipation. It remains unclear to which extent the geometry impacts the amount of kinetic energy that is dissipated. Further studies would be needed to understand whether some particular numerical configurations are more prone to generate heat via turbulent dissipation than others among those employed in CCSN simulations.

\section{Conclusion}
\label{sec:conclusion}
A physical and numerical experiment was conducted to investigate the turbulent regime of neutrino-driven convection in the gain region of a CCSN. Using an idealized model where heating and gravity are localized in the central region of the domain, we addressed several aspects of the competition between convection and downward advection. Parametrized numerical simulations were employed to assess the robustness of linear predictions developed by \citet{foglizzo06} beyond their validity domain. A comparison between 2D and 3D convection enabled us to identify hydrodynamical effects that may result in enhanced heating in global 3D simulations.
This includes a wider mixing region due to a faster rise of buoyant bubbles and stronger viscous heating processes leading to higher entropy values in 3D. In addition, we tested several interpretations of the impact of the dimensionality that are based on hydrodynamical considerations because our setup is suitable to isolate the post-shock hydrodynamics from the complexity of the CCSN modelling. A resolution study has been performed in order to anticipate whether future better resolved global simulations would lead to easier explosions. Our main results and their consequences for CCSN simulations can be summarized as follows.
\newline

1. A set of parametric simulations has been run to characterize the impact of a large amplitude perturbation on a flow in which advection can stabilize convection. We confirm that criteria relying on a balance between buoyancy and gravity or drag force (e.g. \citealt{scheck08, fernandez14}) predict rather accurately the minimal density contrast above which a buoyant bubble is able to rise against advection. However, if the flow has an initial $\chi$ parameter below the linear instability threshold, we observe that convection is eventually suppressed and the flow returns to its initial configuration, regardless of the perturbation amplitude. The damping timescale of the temporary convection increases with larger perturbations and proximity to the linear instability threshold. This indicates that criteria applied to a buoyant bubble do not necessarily predict the development of self-sustained turbulent convection. Our study suggests that the criterion established in the linear regime by \citet{foglizzo06} is also valid in the non linear phase. In situations where convection is linearly stable, non-linear turbulent convection can be triggered by a strong perturbation only if a continuous source of excitation exists, such as SASI or numerical errors.
\newline

2. Our model emphasizes that positive 3D effects may already play an important role in the early non-linear regime of convection. In that phase, 3D bubbles rise faster against the flow than their 2D counterparts. Despite the turbulent drag in 3D, buoyant bubbles resist better against advection in 3D than in 2D. This result is consistent with conclusions drawn from studies on the Rayleigh-Taylor instability with incompressible flows \citep{young01, anuchina04} or Rayleigh-Taylor mixing during the supernova \citep{kane00,hammer10}. These results contradict arguments based on the bubble geometry which are thought to favour 2D explosions. A possible explanation is that the velocity of ascending bubbles is mostly set by their radial extension which does not strongly depend on the dimensionality until the saturation of the instability is reached.
\newline

3. For each of the instability regimes identified in our work, much larger average entropy values are measured in 3D than in 2D (Fig. \ref{fig:timeevol}). This is a clear sign of an additional source of heating acting in 3D. Part of the discrepancies between 2D and 3D may be ascribed to the differences in terms of dynamics. Large scale vortices are a natural outcome of 2D simulations and those tend to stir different entropy phases, which is a less efficient mixing process than small scale mixing. In 2D, the downflows remain unperturbed throughout the gain layer and can channel some matter outside of this region. Turbulent mixing stabilizes a greater fraction of the gain layer in 3D and brings the flow below the marginal stability limit, which is not the case in 2D. Applying a mean-field decomposition on the entropy equation, we identify that turbulent dissipation of kinetic energy could explain the additional heating seen in our 3D simulations. Our results seem to confirm the conjecture of \citet{mabanta18} that this viscous heating process could play a positive role in multidimensional simulations compared to 1D. These even provide a first evidence that entropy generation is stronger in 3D where the turbulent energy cascade favours smaller spatial scales in which kinetic energy is dissipated into heat. 
\newline

4. The impact of resolution is found to be stronger when convection is triggered by a large amplitude perturbation (Fig. \ref{fig:time_reso2}). In such situations, the entropy production rate increases with dimensionality and resolution and the temporary convection is damped on a longer timescale. These results suggest that global 3D simulations with improved numerical resolution might benefit from this mechanism. This could be particularly promising for simulations including precollapse inhomogeneities which are still in their infancy \citep{couch15, mueller17}. This effect is more uncertain when perturbations are numerical artefacts. While this effect would suggest that the discrepancies between Cartesian and spherical grids could widen with finer resolution, this is expected to be counterbalanced by the fact that the numerical noise decreases with resolution. The dynamics seems barely affected by the resolution in the linear instability regime, except for an artificial earlier onset of the instability at low resolution. The level of turbulent dissipation diminishes with higher resolution but seems to be converged with an accuracy of about 20\% for a grid of 64 vertical cells in the gain layer. The parametric study of \citet{radice16} explored resolutions up to six times higher than that and found no clear trend since their highest resolution case quantitatively resembles their lowest one. So far, convergence studies tend to show that explosions are delayed when the numerical resolution is improved \citep{hanke12, abdikamalov15}. It is thus unclear whether higher resolution in CCSN simulations would ease explosions above a certain threshold. It may be that other modeling ingredients would have a stronger impact than resolution. We suggest that the inclusion of large pre-collapse asymmetries could be one of them.
\newline

More realistic studies did not exhibit such a strong dependence of the dimensionality on the average entropy in the gain layer, reaching at most an increase of about 10\% in 3D \citep{dolence13}. However, we note that in that study, 3D explosions occur earlier than in 2D and this is not the case in the work of \citet{hanke12} for which the gap in average entropy values is narrower. It is likely that our setup overestimates the discrepancies between 2D and 3D. A main difference may result from the absence of a stratified cooling layer, in particular in the 2D simulations. Indeed, the vortices are rather expelled out of the gain layer than dissipated inside \citep{murphy11} and the flow cannot adjust itself to a sub-critical one as expected from the study of \citet{fernandez14}. The 3D dynamics might be less affected by the specificities of our setup because the turbulent mixing occurs at smaller spatial scales. Nevertheless, our setup contains the minimal set of ingredients for convection to develop in the presence of advection. Our model correctly mimics the properties of turbulence seen in complex CCSN simulations such as the turbulent energy cascade, the anisotropic turbulence and the higher turbulent pressure in 2D. It enabled us to isolate the hydrodynamics from the complexity of the modelling of CCSNe. We found that several arguments commonly employed in CCSN theory to characterize the greater difficulty to achieve explosions in 3D fail at explaining the discrepancies between 2D in 3D in our simplified model. 

SASI was mostly neglected in this study except to design the initial perturbation. We note that for SASI-dominated cases, idealized models indicate that 3D spiral modes seem more promising to lead to an explosion than 2D sloshing modes because of the larger non-radial kinetic energy stored in spiral modes \citep{fernandez15, kazeroni16}. However, more realistic setups produce delayed explosions in 3D non-rotating SASI-dominated models \citep{melson15b} or even failures \citep{hanke13}. A possibility is that such an additional energy reservoir in 3D SASI models only supports an explosion in conjunction with other physical ingredients such as rotation. Indeed, stellar rotation rates can enhance spiral modes but make explosions harder in 2D \citep{iwakami14b, kazeroni17, summa18}. Newly explored microphysics effects could also create more favourable conditions for the development of SASI \citep{melson15b, bollig17}. In convection-dominated cases, we speculate that turbulent mixing and dissipation might play a similar role as spiral modes for SASI in that they may favour an explosion but their impact alone could be too modest. \citet{melson15a} and \citet{mueller15} pointed out that favourable 3D effects could contribute to the explosion of convection-dominated cases after shock revival. These effects rely on similar physical processes as in our study: forward turbulent cascade, downflow braking due to the growth of the Kelvin-Helmholtz instability in 3D, more efficient turbulent mixing in 3D. Whether these effects are generic is not yet known. 

Finally, a toy model has been used to identify several effects related to the three-dimensional nature of the hydrodynamics that could enhance the heating at the onset of explosion of massive stars. These effects mostly hold for convection-dominated post-shock dynamics, although strong SASI activity can trigger secondary convection. Our idealized model challenges the understanding of convection and the common interpretations of the impact of dimensionality on CCSNe. Our results provide some insights towards a full description of the 3D turbulence generated by neutrino-driven convection. A definitive confirmation of the relevance of the physical processes highlighted in our study will only come from a large set of self-consistent 3D simulations which will become feasible within the next decade. 

\section*{Acknowledgements}
We thank Marc Joos for his help with the analysis tools.
We acknowledge insightful discussions with Bernhard M\"uller, Thomas Janka, Jeremiah Murphy and Luc Dessart.
We are thankful to the referee for helping us improve the manuscript.
This work is part of ANR funded project SN2NS ANR-10-BLAN-0503.
JG acknowledges support from the European Research Council (grant No. 715368 - MagBURST).
This work was granted access to the HPC resources of TGCC/CINES under the allocations t2014047094, x2015047094 and t2016047094 made by GENCI (Grand \'Equipement National de Calcul Intensif) and to the Hydra cluster at Max Planck Computing and Data Facility (MPCDF).



\bibliographystyle{mnras}
\bibliography{convection_bibli} 



\appendix

\section{Mean-field decomposition of the entropy equation}
\label{sec:appendixA}

Using a Favrian decomposition of $S$ and $v_z$ for the left-hand side of the entropy equation (\ref{eq:Sequa}), one obtains:

\begin{equation}
\label{eq:A1}
\begin{aligned}
\frac{\partial\left(\rho S\right)}{\partial t} + \nabla\cdot\left(\rho S v_z\right) = &
\frac{\partial\left(\rho \tilde{S}\right)}{\partial t} + \frac{\partial\left(\rho S^{''}\right)}{\partial t} \\
& + \nabla\cdot\left(\rho \tilde{S} v_z\right) + \nabla\cdot\left(\rho S^{''} \tilde{v_z}\right) + \nabla\cdot\left(\rho S^{''} v_z^{''}\right).
\end{aligned}
\end{equation}

The definition of a Favrian average is such that: $\overline{\rho\tilde{\psi}}=\overline{\rho}\tilde{\psi}=\overline{\rho\psi}$. By definition, the average of a Favrian fluctuation cancels: $\overline{\rho\psi^{''}} = \overline{\rho\left(\psi-\tilde{\psi}\right)} = \overline{\rho\psi}-\overline{\rho\tilde{\psi}}=0$. 
Consequently, the average of the following two terms cancels: $\partial\left(\rho S^{''}\right)/\partial t$ and $\nabla\cdot\left(\rho S^{''} \tilde{v_z}\right)$.

Using the mean-field decomposition (\ref{eq:A1}) to the averaged entropy equation (\ref{eq:Sequa}), one obtains:
\begin{equation}
\label{eq:A2}
 \frac{\partial \bar{\rho}\tilde{S}}{\partial t} + 
 \nabla.\left(\bar{\rho}\tilde{v_z}\tilde{S}\right) + 
 \nabla.\left(\overline{\rho v^{''}_z S^{''}}\right) =
 \overline{\left(\frac{\rho \dot{q}}{T}\right)} +
 \overline{\left(\frac{\rho \epsilon}{T}\right)}.
\end{equation}

\onecolumn

\section{Analytical solution of the neutral stability in the asymptotic limit $K_G\ll1$, $K_H\ll1$, $\M_{\rm up}\ll1$}
\label{sec:appendixB}

\subsection{Differential system}

The evolution of perturbations is described by the same differential system and boundary conditions as in \citet{foglizzo06} considering two cases for the upper boundary condition: either a shock condition, or an acoustic leaking condition. This latter condition expresses that acoustic waves are free to propagate upward and that no acoustic flux nor advected perturbations are injected downward from the region above the heating region.\\
In order to reach an analytical characterization of the heating threshold for a convective instability, we chose a regime where the flow is approximately uniform, thus allowing for an analytical integration of the differential system. This is achieved in the limit of small gravity $(K_G\ll1$) and small external heating ($K_H\ll1$). In this regime, having comparable advection and buoyancy timescales (i.e. finite non zero $\chi$ parameter) requires a small enough mach number ($\M_{\rm up}^2\propto K_HK_G\ll1$). This asymptotic differential system is obtained by introducing some new functions ($\delta \hat F,\delta \hat h$) defined by
\begin{eqnarray}
\delta \hat F&\equiv& {1\over \M}\left({\delta f\over c^2}-{\delta S\over\gamma}\right),\\
{\delta \hat h}&\equiv&\M\left(\delta h+{\gamma-1\over\gamma}\delta S\right).
\end{eqnarray}
where $\delta f \equiv v\delta v_z + 2c\delta c/(\gamma-1)$ and $\delta h \equiv \delta v_z/v + \delta \rho/\rho$. Conversely,
\begin{eqnarray}
{\delta \rho\over \rho}&=&{1\over 1-\Mc}\left(\M(\delta \hat F-\delta \hat h)-{\gamma-1\over\gamma}(1-\M^2){\delta S}\right),\\
{\delta v_z\over v}&=&{1\over 1-\Mc}{1\over\M}\left({\delta \hat h}-\M^2{\delta \hat F}\right),\\
{\delta c^2\over c^2}&=&{\gamma-1\over 1-\Mc}\left(\M({\delta \hat F}-\delta\hat h)+(1-\M^2) {\delta S\over\gamma}\right).
\end{eqnarray}
The differential system satisfied by ($\delta \hat F,\delta \hat h, \delta S,\delta v_x$) for neutral stability ($\omega=0$) is the following
\begin{eqnarray}
{\p \delta \hat F\over\p z}&=&
{\gamma-1\over 1-\Mc}{\nabla\Phi\over \M c^2}\left({\delta S\over\gamma}+\M\delta \hat F\right)
-{1\over\M}{\p \M\over\p z}\delta \hat F
-\left(
\delta\hat h
-{\gamma-1\over \gamma}\M\delta S
-\gamma\Mc\delta \hat F
\right)
{1\over 1-\M^2}{\gamma-1\over\gamma}\nabla S
, \\
{\p\delta \hat h\over\p z}&=&
k_x{i\delta v_x\over c}
+{\gamma-1\over\gamma}\M\delta\left({{\cal L}\over p v}\right)
+{1\over\M}{\p \M\over\p z}\delta \hat h,\\
{\p\delta S\over\p z}&=&\delta\left({{\cal L}\over p v}\right),\\
{\p i\delta v_x\over\p z}&=&-{k_xc}\delta \hat F
+k_x{c\over 1-\Mc}\left({\delta \hat h}-\M^2{\delta \hat F}\right),
\end{eqnarray}
When $\M\ll1$, $H|\nabla S|\ll1$ and $H\nabla\Phi/c^2\ll1$, the flow is approximately uniform and the leading terms of the differential system are independent of the detailed form of the heating function. For analytical simplicity, we neglect the linear transition ramp used to smoothen the gravity profile and the heating function ($\Psi(|z|<H)=1$) such that $K_H=-H\nabla S$ and $K_G=H\nabla \Phi/c_{\rm up}^2$. Assuming $K_H\propto\M_{\rm up}$ and $K_G\propto\M_{\rm up}$ in the limit where $\M_{\rm up}\ll1$, the leading coefficients of the differential system are constants of order unity:
\begin{eqnarray}
{\p \delta \hat F\over\p z}&=&
{\gamma-1\over \gamma}{K_G\over H\M_{\rm up} }{\delta S}, \label{eq:differential_delta_f_limit}\\
{\p\delta \hat h\over\p z}&=&
k_x{i\delta v_x\over c_{\rm up}},\\
{\p\delta S\over\p z}&=&{K_H\over H \M_{\rm up}}\delta\hat h,\\
{\p\over\p z}{i\delta v_x\over c_{\rm up}}&=&{k_x}({\delta \hat h}-\delta \hat F). \label{eq:differential_delta_v_limit}
\end{eqnarray}
The eigenfunctions ($\delta \hat F, \delta \hat h, \delta S, i\delta v_x/c_{\rm up}$) are a linear combination of four functions $\exp(\beta_j\chi k_x z)$ with respective coefficients $(f_j, h_j, s_j,v_j)_{j=1,4}$. An important property of the solution is that the parameters $K_H$, $K_G$, $\M_{\rm up}$ play a role only through the combination $\chi$ defined by
\begin{eqnarray}
\label{eq:chi_params}
\chi^2&\equiv& 4{\gamma-1\over\gamma}{K_HK_G\over\M_{\rm up}^2},\\
&=&\left({\omega_{\rm BV} 2H\over v_{\rm up}}\right)^2.
\end{eqnarray}
$\beta_{j=1,4}$ are the four roots of the equation $\beta^4-\beta^2+1/4\alpha^2=0$ (deduced from the determinant of the system of equations~\ref{eq:differential_delta_f_limit}-\ref{eq:differential_delta_v_limit}) with
\begin{eqnarray}
\alpha&\equiv&{H {k_x}\over \chi}.
\end{eqnarray}
If $0\le\alpha\le1$, we define the angle $\phi$ with $0\le\phi\le\pi/2$ such that $\alpha\equiv \cos\phi$:
\begin{eqnarray}
\beta_1&\equiv&{\e^{ i{\phi\over2}}\over(2\cos\phi)^{1\over2}},\\
\beta_2&\equiv&\bar{\beta}_1={\e^{-i{\phi\over2}}\over(2\cos\phi)^{1\over2}},\\
\beta_3&\equiv&-\beta_1,\\
\beta_4&\equiv&-\beta_2.
\end{eqnarray}

\subsection{Boundary conditions}

\subsubsection{Decomposition into advected and acoustic waves}

In the uniform semi-infinite regions above and below the region of heating, the perturbation is decomposed into 4 components: entropic, vortical, and two acoustic waves propagating with the flow (i.e. downward, subscript $+$) or against the flow (i.e. upward, subscript $-$)
\begin{eqnarray}
\delta \hat F=\delta \hat F_S+\delta \hat F_w+\delta \hat F_++\delta \hat F_-.
\end{eqnarray}
The identification of acoustic waves propagating  can be adapted from Eqs.~(C16-C21) in \citet{foglizzo06}. The pressure perturbation ($\delta \hat F_\pm,\delta \hat h_\pm$) with a vertical wavenumber $k_z^\pm$ is characterized by the absence of perturbation of entropy ($\delta S=0$) and vorticity ($k_xc^2\delta \hat F_\pm=0$). 
With $k_z^\pm= \mp ik_x$ in the low Mach limit, the pressure profile of a purely growing mode $\exp(ik_z^-z)$ is evanescent upward while $\exp(ik_z^+z)$ is evanescent downward. Using this formula for the expression of $\delta \hat h_\pm$,
\begin{eqnarray}
\delta \hat F_\pm&=&0,\\
\delta \hat h_\pm&=&\pm i{\delta v_x^\pm\over c}.
\end{eqnarray}

The advected perturbation containing entropy and vorticity perturbations is purely advected ($\omega = k_z v$). From equations~\ref{eq:differential_delta_f_limit}-\ref{eq:differential_delta_v_limit} with $k_z=\omega/v=0$ for neutral stability, we deduce:

\begin{eqnarray}
\delta v_{x \, \rm adv} &=& 0, \\
\delta \hat F_{\rm adv}&=& \delta \hat h_{\rm adv}.
\end{eqnarray}

\subsubsection{Leaking lower boundary}

The lower boundary condition $\delta \hat h_-^{\rm d}=0$ is expressed in the low Mach limit:
\begin{eqnarray}
\delta \hat F_{\rm d}
-
\delta \hat h_{\rm d}
+{i\delta v_x^{\rm d}\over c_{\rm d}}
=0.
\end{eqnarray} 

\subsubsection{Leaking upper boundary condition without a shock}

The upper boundary condition corresponds to a pure acoustic wave propagating against the flow: $\delta S=0$, $\delta F_{\rm adv}=0$, $\delta F_+=0$. In the low Mach limit,
\begin{eqnarray}
\delta S_{\rm up}&=&0,\label{kn0S}\\
\delta \hat F_{\rm up}&=&0,\label{kn0F}\\
\delta \hat h_{\rm up}&=&-{i\delta v_x^{\rm up}\over c_{\rm up}}.\label{kn0h}
\end{eqnarray} 

\subsubsection{Upper boundary condition with a shock}

The boundary condition for a perturbed shock with dissociation established by \citet{foglizzo06} (Eqs. 28-30 and C14), are written using the functions $\delta\hat F,\delta\hat h$, at marginal stability. In the uniform limit where $\M_{\rm sh}\ll1$, $H\nabla S\ll1$ and $H\nabla\Phi/c_{\rm sh}^2\ll1$,
\begin{eqnarray}
\delta \hat h_{\rm sh}&=&0,\\
\delta S_{\rm sh}&=&0,\\
\delta\hat F_{\rm sh}&=&0.
\end{eqnarray}

\subsection{Analytical formulation of the criterion for marginal stability}

\subsubsection{Marginal stability without a shock}

\begin{figure}
\centering
	\includegraphics[width=0.48\columnwidth]{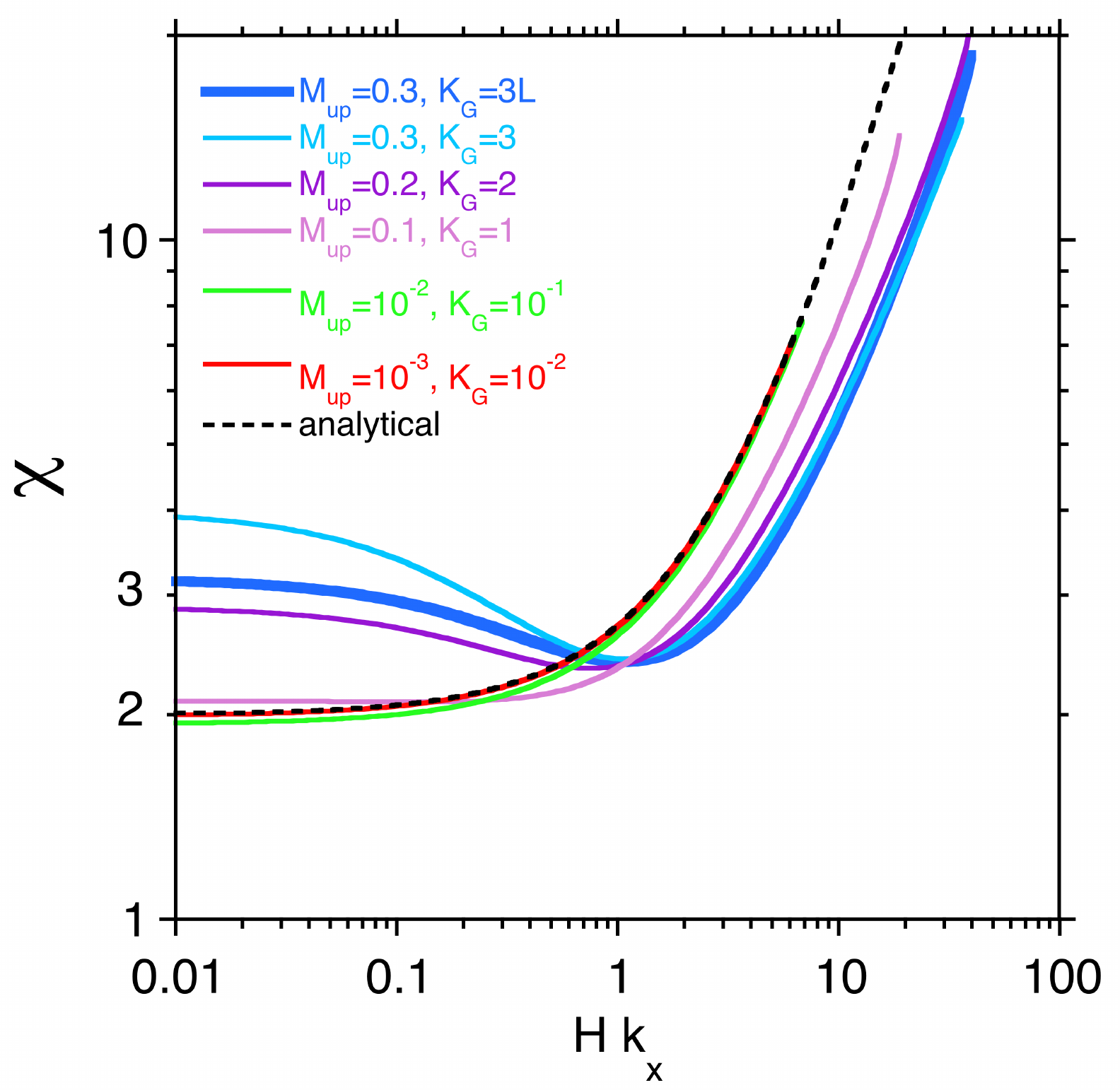}
	\includegraphics[width=0.48\columnwidth]{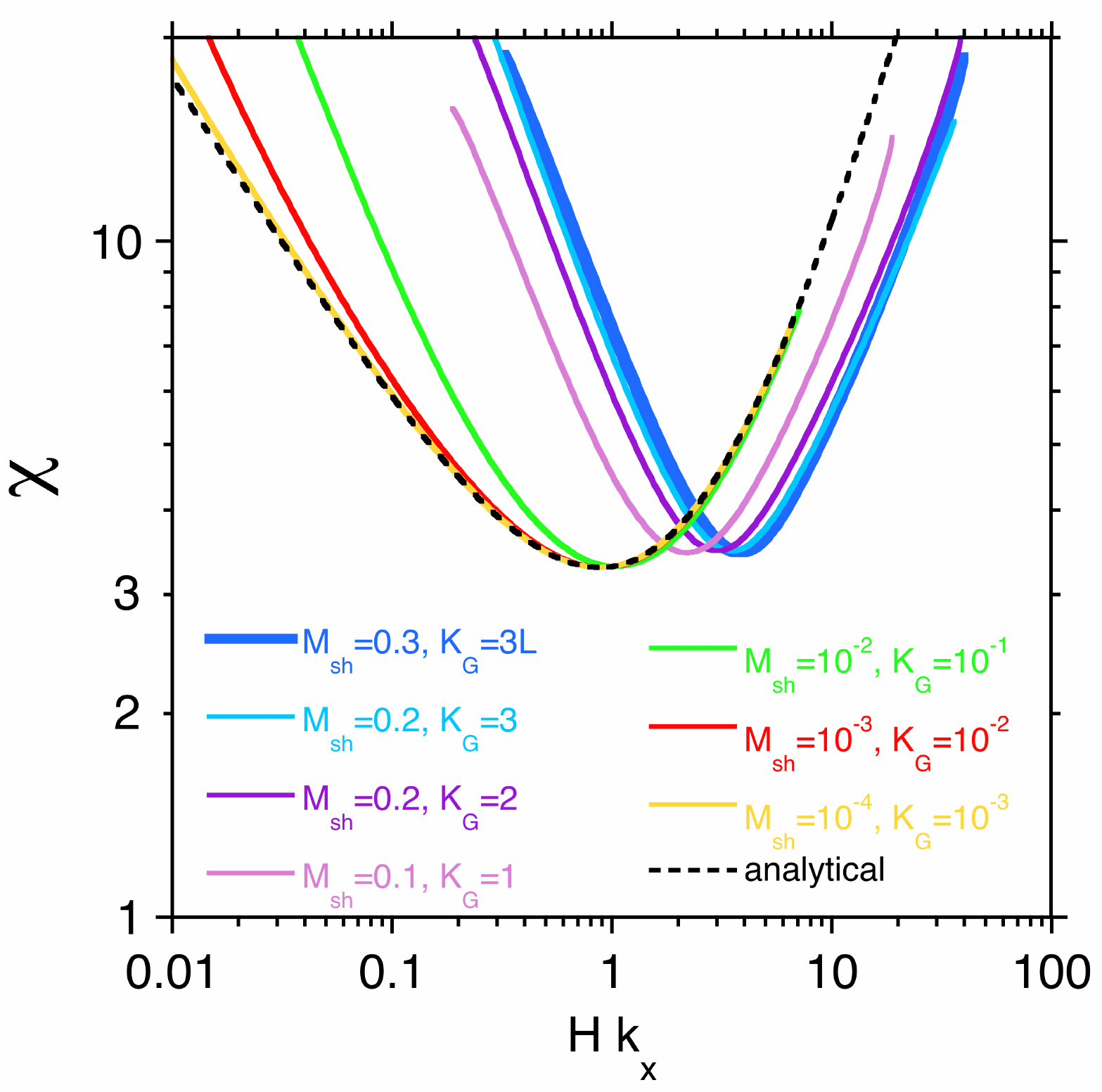}
    \caption{Convergence of the instability threshold $\chi(Hk_x)$ towards the analytical expression (dashed line) when $K_G\ll1$, either in a subsonic flow (${\cal M}_{\rm up}\ll1$, left panel) or in a postshock flow (${\cal M}_{\rm sh}\ll1$, right panel). The curves have been calculated without a transition function $\Psi(|z|<1)=1$, except for the thick curve noted $K_G=3L$ which shows the incidence of a linear ramp $\Psi(|z|<0.5)=1$. The analytical solutions correspond to Eqs. (\ref{eq:chi_noshock}) and (\ref{eq:chi_shock}).}
    \label{fig:convergence}
\end{figure}

With $z_{\rm d}=-H$, the differential system is translated into linear relations between the coefficients $(f_j, h_j, s_j,v_j)_{j=1,4}$:
\begin{eqnarray}
f_j&=& {1\over4\alpha^2\beta_j^2} h_j,\\
s_j&=&{K_H\over\M\alpha\beta_j\chi}h_j,\\
v_j&=& {\beta_j}h_j.
\end{eqnarray}
The boundary conditions are translated into:
\begin{eqnarray}
\sum_{j=1}^4 s_j&=&0,\\
\sum_{j=1}^4 f_j&=&0,\\
\sum_{j=1}^4 (v_j+h_j)&=&0,\\
\sum_{j=1}^4 (f_j-h_j+v_j)\e^{-2\chi\alpha\beta_j}&=&0.
\end{eqnarray}
The value of $\alpha(\chi)$ is set by the vanishing determinant of the linear system satisfied by $(h_1, h_2, h_3, h_4)$:
\begin{eqnarray}
\begin{vmatrix}
{\beta_2}&
{\beta_1}&
-{\beta_2}&
-{\beta_1}\\
{\beta_2^2}&
{\beta_1^2}&
{\beta_2^2}&
{\beta_1^2}\\
1+\beta_1&
1+\beta_2&
1-\beta_1&
1-\beta_2\\
\left(\beta^2_1-\beta_1\right)\e^{-2\chi\alpha\beta_1}&
\left(\beta^2_2-\beta_2\right)\e^{-2\chi\alpha\beta_2}&
\left(\beta^2_1+\beta_1\right)\e^{2\chi\alpha\beta_1}&
\left(\beta^2_2+\beta_2\right)\e^{2\chi\alpha\beta_2}
\end{vmatrix}
=0,
\end{eqnarray}
where we have used $1/\beta_1=(2\cos\phi)\beta_2$. The determinant $\Delta_{\rm leak}$ can be factorized as follows
\begin{eqnarray}
\Delta_{\rm leak}&=&{8\over 4\alpha^2}(\beta_2^2-\beta_1^2)(Y_1-Y_2)\left(
1-{1\over Y_1Y_2}
\right),\\
Y_1&\equiv&4\alpha^2(\beta^2_1+\beta_1)^2\e^{2\chi\alpha\beta_1},\\
Y_2&\equiv&4\alpha^2(\beta^2_2+\beta_2)^2\e^{2\chi\alpha\beta_2}.
\end{eqnarray}
It can be shown that the only vanishing factor in the factorized expression of $\Delta_{\rm leak}$ is $Y_1-Y_2$. For $\alpha(\chi)<1$ the analytical relation between the value of $\chi$ and $Hk_x$ is thus reduced to:
\begin{eqnarray}
\label{eq:chi_noshock}
\chi&=&{2^{1\over2}\over\cos^{1\over2}\phi\sin{\phi\over2}}
\arctan{\cos^{1\over2}\phi
\over
2^{1\over2}\sin{\phi\over2}
},\\
\cos\phi&\equiv&{Hk_x\over\chi}.
\end{eqnarray}
The solution $Hk_x=\chi$ when $\chi\gg1$ corresponds to a stabilization a short wavelength when
\begin{eqnarray}
k_x={2\over |v|}\omega_{\rm BV}(\chi\gg1).
\end{eqnarray}
It agrees within a factor 2 with the intuitive expectation based on the evanescent vertical profile of the Rayleigh Taylor instability ($\exp (\omega_i -k_x|v|)t$) \citep{guilet10}. The limit $\phi=\pi/2$ corresponds to the threshold of stability when $k_x\to 0$ and $\chi\to 2$. The flow is linearly stable for $\chi<2$ (Fig. \ref{fig:convergence}, left panel).

\subsection{Marginal stability with a shock}

The boundary conditions are translated into:
\begin{eqnarray}
\sum_{j=1}^4 f_j&=&0,\\
\sum_{j=1}^4 s_j&=&0,\\
\sum_{j=1}^4 h_j&=&0,\\
\sum_{j=1}^4 (f_j-h_j+v_j)\e^{-2\chi\alpha\beta_j}&=&0,
\end{eqnarray}
with
\begin{eqnarray}
f_j&=& {1\over4\alpha^2\beta_j^3} v_j
,\\
s_j&=&{K_H\over\M\alpha\beta_j^2\chi}v_j,\\
h_j&=& {1\over\beta_j}v_j.
\end{eqnarray}
The value of $\alpha(\chi)$ is set by the vanishing determinant of the linear system satisfied by $(v_1, v_2, v_3, v_4)$:
\begin{eqnarray}
\begin{vmatrix}
\beta_2&
\beta_1&
-\beta_2&
-\beta_1\\
\beta_2^2&
\beta_1^2&
\beta_2^2&
\beta_1^2\\
\beta_2^3&
\beta_1^3&
-\beta_2^3&
-\beta_1^3\\
(1-\beta_1)\e^{-2\chi\alpha\beta_1}&
(1-\beta_2)\e^{-2\chi\alpha\beta_2}&
(1+\beta_1)\e^{2\chi\alpha\beta_1}&
(1+\beta_2)\e^{2\chi\alpha\beta_2}
\end{vmatrix}
=0
\end{eqnarray}
Defining
\begin{eqnarray}
Y_3&\equiv&
(1+\beta_1)\e^{2\chi\alpha\beta_1}
+(1-\beta_1)\e^{-2\chi\alpha\beta_1},\\
Y_4&\equiv&
(1+\beta_2)\e^{2\chi\alpha\beta_2}
+(1-\beta_2)\e^{-2\chi\alpha\beta_2},
\end{eqnarray}
the determinant $\Delta_{\rm shock}$ is factorized as follows:
\begin{eqnarray}
\Delta_{\rm shock}=8\beta_1\beta_2(\beta_1^2-\beta_2^2)(\beta_1^2Y_3-\beta_2^2Y_4)
\end{eqnarray}
For $\alpha=Hk_x/\chi<1$ we transform the equation $\beta_1^2Y_3-\beta_2^2Y_4=0$ into an implicit equation relating $\chi$ and $Hk_x$:
\begin{eqnarray}
\cos\phi&\equiv&{Hk_x\over \chi},\\
X_c&\equiv&\chi\times(2\cos\phi)^{1\over2}\cos{\phi\over2},\\
X_s&\equiv&\chi\times(2\cos\phi)^{1\over2}\sin{\phi\over2},\\
\label{eq:chi_shock}
(2\cos\phi)^{1\over2}\left(\cos\phi
\sin X_s
\tanh X_c
+\sin\phi\cos X_s\right)
&+&\cos{3\phi\over2}
\sin X_s
+\sin{3\phi\over2}\cos X_s
\tanh X_c=0.
\end{eqnarray}
This equation has two solutions $k_{\rm min}$, $k_{\rm max}$ when $\chi>\chicrit\simeq3.291$.\\
Using a Taylor expansion of $\tanh X_c$ and $\tan X_s$ for $\chi\gg1$, $Hk_{\rm min}\ll1$, $\phi\sim\pi/2$, $\alpha\ll 1$, we obtain
\begin{eqnarray}
Hk_{\rm min}&=&{3\over\chi^2}.
\end{eqnarray}
If $\chi\gg1$ and $Hk_{\rm max}\gg1$, $\phi\ll1$, $\alpha=1$,
\begin{eqnarray}
Hk_{\rm max}&=&{\chi}.
\end{eqnarray}
The threshold for global instability corresponds to $H k_{\rm min}=H k_{\rm max}\sim1$ and $\chicrit\simeq3.291$ (Fig. \ref{fig:convergence}, right panel).


\bsp	
\label{lastpage}
\end{document}